\documentclass[a4paper,fleqn,usenatbib]{mnras}

\usepackage{newtxtext,newtxmath}

\usepackage[T1]{fontenc}
\usepackage{ae,aecompl}

\usepackage{graphicx}	
\usepackage{amsmath}	
\usepackage{amssymb}	
\usepackage{multicol}        
\usepackage{bm}		
\usepackage{pdflscape}	
\usepackage{longtable}

\newcommand{\kms}{\,km\,s$^{-1}$} 
\newcommand{\niceurl}[1]{\href{#1}{#1}}

\newcommand{\oii}{$[\textnormal{O}_\textsc{II}]$} 
\newcommand{\zspec}{z_{\rm spec}} 

\graphicspath{{plots/}}

\usepackage[T1]{fontenc}
\usepackage{ae,aecompl}

\usepackage{newtxtext,newtxmath}
\usepackage{multirow}
\usepackage{tablefootnote}

\usepackage[dvipsnames]{xcolor}

\title[SDSS/eBOSS DR16 ELG LSS and isotropic BAO]{\textit{The completed SDSS-IV extended Baryon Oscillation Spectroscopic Survey: Large-scale Structure Catalogues and Measurement of the isotropic BAO between redshift 0.6 and 1.1 for the Emission Line Galaxy Sample}}

\author[A. Raichoor et al.]{\parbox{\textwidth}{
Anand Raichoor$^{1\thanks{e-mail: \texttt{\href{mailto:anand.raichoor@epfl.ch}{anand.raichoor@epfl.ch}}}}$,
Arnaud de Mattia$^{2}$,
Ashley J. Ross$^{3}$,
Cheng Zhao$^{1}$,
Shadab Alam$^{4}$,
Santiago Avila$^{5,6}$,
Julian Bautista$^{7}$,
Jonathan Brinkmann$^{8}$,
Joel R. Brownstein$^{9}$,
Etienne Burtin$^{2}$,
Michael J. Chapman$^{10,11}$,
Chia-Hsun Chuang$^{12}$,
Johan Comparat$^{13}$,
Kyle S. Dawson$^{9}$,
Arjun Dey$^{14}$,
H\'elion du Mas des Bourboux$^{9}$,
Jack Elvin-Poole$^{3}$,
Violeta Gonzalez-Perez$^{15,7}$,
Claudio Gorgoni$^{1}$,
Jean-Paul Kneib$^{1,16}$,
Hui Kong$^{3}$,
Dustin Lang$^{17,11}$,
John Moustakas$^{18}$,
Adam D. Myers$^{19}$,
Eva-Maria M\"uller$^{20}$,
Seshadri Nadathur$^{7}$,
Jeffrey A. Newman$^{21}$,
Will J. Percival$^{10,11,17}$,
Mehdi Rezaie$^{22}$,
Graziano Rossi$^{23}$,
Vanina Ruhlmann-Kleider$^{2}$,
David J. Schlegel$^{24}$,
Donald P. Schneider$^{25,26}$,
Hee-Jong Seo$^{22}$,
Am\'elie Tamone$^{1}$,
Jeremy L. Tinker$^{27}$,
Rita Tojeiro$^{28}$,
M. Vivek$^{25,29}$,
Christophe Y\`eche$^{2,24}$,
Gong-Bo Zhao$^{30,31,7}$
 } \vspace*{4pt} \\ 
\small $^{1}$ Institute of Physics, Laboratory of Astrophysics, Ecole Polytechnique Fédérale de Lausanne (EPFL), Observatoire de Sauverny, 1290 Versoix, Switzerland\vspace*{-2pt} \\ 
\small $^{2}$ IRFU, CEA, Universit\'e Paris-Saclay, F-91191 Gif-sur-Yvette, France\vspace*{-2pt} \\ 
\small $^{3}$ Center for Cosmology and AstroParticle Physics, The Ohio State University, Columbus, OH 43212\vspace*{-2pt} \\ 
\small $^{4}$ Institute for Astronomy, University of Edinburgh, Royal Observatory, Blackford Hill, Edinburgh, EH9 3HJ , UK\vspace*{-2pt} \\ 
\small $^{5}$ Universidad Aut\'onoma de Madrid, 28049, Madrid, Spain\vspace*{-2pt} \\ 
\small $^{6}$ Instituto de Fisica Teorica UAM/CSIC, Universidad Autonoma de Madrid, 28049 Madrid, Spain\vspace*{-2pt} \\ 
\small $^{7}$ Institute of Cosmology \& Gravitation, University of Portsmouth, Dennis Sciama Building, Burnaby Road, Portsmouth PO1 3FX, UK\vspace*{-2pt} \\ 
\small $^{8}$ Apache Point Observatory and New Mexico State University, P.O. Box 59, Sunspot, NM 88349\vspace*{-2pt} \\ 
\small $^{9}$ University of Utah, Department of Physics and Astronomy, 115 S 1400 E, Salt Lake City, UT 84112, USA\vspace*{-2pt} \\ 
\small $^{10}$ Waterloo Centre for Astrophysics, University of Waterloo, Waterloo, ON~N2L~3G1, Canada\vspace*{-2pt} \\ 
\small $^{11}$ Department of Physics and Astronomy, University of Waterloo, 200 University Ave W, Waterloo, ON N2L 3G1, Canada\vspace*{-2pt} \\ 
\small $^{12}$ Kavli Institute for Particle Astrophysics and Cosmology, Stanford University, 452 Lomita Mall, Stanford, CA 94305, USA\vspace*{-2pt} \\ 
\small $^{13}$ Max-Planck-Institut f\"{u}r extraterrestrische Physik (MPE), Giessenbachstrasse 1, D-85748 Garching bei M\"unchen, Germany\vspace*{-2pt} \\ 
\small $^{14}$ NSF's National Optical-Infrared Astronomy Research Laboratory, 950 N. Cherry Ave, Tucson, AZ 85719 USA\vspace*{-2pt} \\ 
\small $^{15}$ Astrophysics Research Institute, Liverpool John Moores University, 146 Brownlow Hill, Liverpool L3 5RF, UK\vspace*{-2pt} \\ 
\small $^{16}$ Aix Marseille Univ, CNRS, CNES, LAM, Marseille, France\vspace*{-2pt} \\ 
\small $^{17}$ Perimeter Institute, Waterloo, ON~N2L~2Y5, Canada\vspace*{-2pt} \\ 
\small $^{18}$ Department of Physics \& Astronomy, Siena College, 515 Loudon Road, Loudonville, NY 12211, USA\vspace*{-2pt} \\ 
\small $^{19}$ Department of Physics and Astronomy, University of Wyoming, Laramie, WY 82071, USA\vspace*{-2pt} \\ 
\small $^{20}$ Department of Physics, University of Oxford, Denys Wilkinson Building, Keble Road, Oxford OX1 3RH, U.K\vspace*{-2pt} \\ 
\small $^{21}$ University of Pittsburgh and PITT PACC, Pittsburgh, PA 15260\vspace*{-2pt} \\ 
\small $^{22}$ Department of Physics and Astronomy, Ohio University, Clippinger Labs, Athens, OH 45701, USA\vspace*{-2pt} \\ 
\small $^{23}$ Department of Physics and Astronomy, Sejong University, Seoul, 143-747, Korea\vspace*{-2pt} \\ 
\small $^{24}$ Lawrence Berkeley National Laboratory, Berkeley, CA~94720, USA\vspace*{-2pt} \\ 
\small $^{25}$ Department of Astronomy and Astrophysics, The Pennsylvania State University, University Park, PA 16802\vspace*{-2pt} \\ 
\small $^{26}$ Institute for Gravitation and the Cosmos, The Pennsylvania State University, University Park, PA 16802\vspace*{-2pt} \\ 
\small $^{27}$ Center for Cosmology and Particle Physics, Department of Physics, New York University, 726 Broadway, Room 1005, New York, NY 10003, USA\vspace*{-2pt} \\ 
\small $^{28}$ School of Physics and Astronomy, University of St Andrews, North Haugh, St Andrews, KY16 9SS, UK\vspace*{-2pt} \\ 
\small $^{29}$ Indian Institute of Astrophysics, Koramangala, Bangalore 560034, India\vspace*{-2pt} \\ 
\small $^{30}$ National Astronomical Observatories of China, Chinese Academy of Sciences, 20A Datun Road, Chaoyang District, Beijing 100012, China\vspace*{-2pt} \\ 
\small $^{31}$ University of Chinese Academy of Sciences, Beijing, 100049, China\vspace*{-2pt} \\ 
}

\date{Last updated xxx; in original form xxx}

\pubyear{2018}

\begin{document}
\label{firstpage}
\pagerange{\pageref{firstpage}--\pageref{lastpage}}
\maketitle

\begin{abstract}
We present the Emission Line Galaxy (ELG) sample of the extended Baryon Oscillation Spectroscopic Survey (eBOSS) from the Sloan Digital Sky Survey IV Data Release 16 (DR16).
After describing the observations and redshift measurement for the 269,243 observed ELG spectra over 1170 deg$^2$, we present the large-scale structure catalogues, which are used for the cosmological analysis.
These catalogues contain 173,736 reliable spectroscopic redshifts between 0.6 and 1.1, along with the associated random catalogues quantifying the extent of observations, and the appropriate weights to correct for non-cosmological fluctuations.
We perform a spherically averaged baryon acoustic oscillations (BAO) measurement in configuration space, with density field reconstruction: the data 2-point correlation function shows a feature consistent with that of the BAO, providing a 3.2-percent measurement of the spherically averaged BAO distance $D_V(z_{\rm eff})/r_{\rm drag} = 18.23\pm 0.58$ at the effective redshift $z_{\rm eff}=0.845$.
\end{abstract}





\section{Introduction} \label{sec:intro}

The acceleration of the expansion of the Universe discovered about twenty years ago \citep{riess98,perlmutter99} set a key milestone in cosmology history: current observations can be accounted for with the $\Lambda$CDM standard model, but at the cost of introducing a dark energy component, making up today $\sim$70 percent of the energy content of the Universe.
Around the same time, the SDSS collaboration \citep{york00} initiated spectroscopic observations to study large-scale structures, which allows one to constrain the geometry of the Universe with the Baryonic Acoustic Oscillations \citep[BAO,][]{eisenstein98} and the growth of structures with redshift space distortion \citep[RSD,][]{kaiser87}.

Since then, the SDSS has become a key experiment for the BAO , one of the most powerful cosmological probes \citep[see ][for a review]{weinberg13a}.
The SDSS first measured the distance-redshift relation with 5 percent precision at $z=0.35$ \citep{eisenstein05} from 45,000 Luminous Red Galaxies \citep[LRGs,][]{eisenstein01}.
It was the first BAO detection along with the 2dF Galaxy Redshift Survey \citep{colless03,cole05}.
The BOSS survey \citep[2008--2014,][]{dawson13} from the SDSS-III \citep{eisenstein11} then massively observed 1.5 million LRGs and 160,000 quasars (QSOs), leading to a state-of-the-art 1--2 percent precision measurement of the cosmological distance scale for redshifts $z<0.6$ \citep{alam17} and $z=2.5$ \citep{delubac15,bautista17}.
The Extended Baryon Oscillation Spectroscopic Survey \citep[eBOSS, 2014--2020,][]{dawson16} of the SDSS-IV \citep{blanton17} observed nearly one million objects to complement the BOSS survey in the $0.6<z<2.2$ redshift range.
eBOSS observed LRGs at $0.6<z<1.0$ \citep{prakash16}, Emission Line Galaxies at $0.6<z<1.1$ \citep[ELGs,][]{raichoor17}, and QSOs at $0.9<z<3.5$ \citep{myers15,palanque-delabrouille16}.

We present in this paper the eBOSS/ELG spectroscopic observations from the final release from SDSS-IV, DR16 \citep{ahumada19}, along with the construction of the large-scale structure (LSS) catalogues, and the spherically-averaged BAO measurement from those.
The LSS catalogues are also used in \citet{de-mattia20} and \citet{tamone20} to analyse the ELG anistropic clustering.
ELGs are star-forming galaxies with strong emission lines -- noticeably the \oii~ doublet emitted at ($\lambda$3727, $\lambda$3729 \AA), allowing a spectroscopic redshift ($\zspec$) measurement in a reasonable amount of exposure time, as there is no need to significantly detect the continuum.
This observational feature, combined with their abundance at $z \sim 0.5$--2 due to the high star-formation density of the Universe then \citep[e.g.,][]{lilly96,madau98,madau14}, make them a promising tracer for large-scale structures surveys.
The WiggleZ experiment \citep[2006--2011,][]{drinkwater10} was the first survey to use ELGs.
Now eBOSS paves the way for the next generation LSS surveys, which will heavily rely on the ELGs in the $0.5 \lesssim z \lesssim 2$ range, as  
PFS\footnote{Prime Focus Spectrograph: \niceurl{http://sumire.ipmu.jp/en/2652/}} \citep{sugai12, takada14},
DESI\footnote{Dark Energy Spectroscopic Instrument: \niceurl{http://desi.lbl.gov/cdr/}} \citep{desi-collaboration16a,desi-collaboration16b},
4MOST\footnote{4-meter Multi-Object Spectroscopic Telescope: \niceurl{https://www.4most.eu/}} \citep{de-jong14},
 \textit{Euclid} \citep{laureijs11}, and \textit{WFIRST}\footnote{Wide-Field Infrared Survey Telescope: \niceurl{https://wfirst.gsfc.nasa.gov/}} \citep{dore18}.
Indeed, this eBOSS/ELG sample has already been used for several analyses, which strengthen our understanding of ELGs at $z \sim 1$: exploring their physical content \citep{gao18,huang19}, their dark matter halos properties \citep{gonzalez-perez18,guo19,gonzalez-perez20} and alternative methods to improve the removal of systematics in their clustering \citep{rezaie19,kong20}.

This paper is part of a series of papers presenting the final eBOSS DR16 data and cosmological results.
The LRG and QSO LSS catalogues are presented in \citet{ross20}; the QSOs LSS catalogues use the DR16 QSO catalogues presented in \citet{lyke20}.
The N-body mocks, along with mock challenges done to validate the eBOSS analysis, are presented in \citet[][LRGs]{rossi20}, \citet[][ELGs]{alam20,avila20}, and \citet[][QSOs]{smith20}.
The approximate mocks are presented in \citet[][EZmocks]{zhao20} and \citet[][QPM-GLAM]{lin20}.
The anisotropic clustering analyses are presented in configuration space in \citet[][LRGs]{bautista20}, \citet[][ELGs]{tamone20}, \citet[][ELGs and LRGs]{wang20},  \citet[][QSOs]{hou20}, and in Fourier space in \citet[][LRGs]{gil-marin20}, \citet[][ELGs]{de-mattia20}, \citet[][ELGs and LRGs]{zhao20a}, and \citet[][QSOs]{neveux20}.
The Ly-$\alpha$ auto- and cross-correlation are presented in \citet{dumasdesbourboux20}.
Lastly, the cosmological implication of the full eBOSS sample is presented in \citet{eboss20}.
A summary of all SDSS BAO and RSD measurements with accompanying legacy figures, along with he full cosmological interpretation of these measurements is available online\footnote{\niceurl{
https://www.sdss.org/science/final-bao-and-rsd-measurements/}, \niceurl{https://www.sdss.org/science/cosmology-results-from-eboss/}}.

The paper is organized as follows.
Section \ref{sec:data} briefly summarises the target selection and presents the spectroscopic observations and the $\zspec$ measurement.
The building of the LSS catalogues is detailed in Section \ref{sec:lss}, including the random catalogue construction, the angular veto masking, and the definition of the weights to correct for non-cosmological fluctuations in the data.
The mock catalogues used for the spherically averaged BAO analysis are introduced in Section \ref{sec:mocks}, and the spherically averaged BAO analysis in configuration space is presented in Section \ref{sec:bao}.
We conclude in \mbox{Section \ref{sec:conclusion}}.

\section{Data} \label{sec:data}

We describe in this Section the target selection, the spectroscopic observations and the spectroscopic redshift ($\zspec$) estimation of the eBOSS/ELG sample.

\subsection{Imaging and target selection}
The ELG target selection is extensively described in \citet{raichoor17}, to which we refer the reader for more details.

Targets are selected using the DECaLS part of the Legacy Imaging Surveys\footnote{\niceurl{http://legacysurvey.org/}} \citep{dey19} $grz$ photometry, which provides the imaging for the DESI target selection.
In detail, the DECaLS program is a consistent processing of public imaging taken with the Dark Energy Camera \citep[DECam][]{flaugher15}, mostly coming from the DECaLS survey (co-PIs: A. Dey and D.J. Schlegel; NOAO Proposal \# 2014B-0404) and the DES\footnote{\niceurl{http://www.darkenergysurvey.org}}  (PI: J. Frieman; NOAO Proposal \# 2012B-0001).
\citet{comparat16a} and \citet{raichoor16} demonstrated that DECaLS permits a better target selection in terms of higher redshift and density, than the SDSS imaging.
The footprint is divided in two parts (see Figure \ref{fig:tiling}): $\sim$620 deg$^{2}$ in the Fat Stripe 82 in the South Galactic Cap (SGC) at $-43^\circ$<R.A.<45$^\circ$ and $-5^\circ$<Dec.<5$^\circ$, covered by the DES and $\sim$550 deg$^{2}$ in the North Galactic Cap (NGC) at 126$^\circ$<R.A.<166$^\circ$ and 13.8$^\circ$<Dec.<32.5$^\circ$, covered by the DECaLS survey.
The DES imaging we use in the SGC is $\sim$0.5 mag deeper than the DECaLS imaging used in the NGC.

The target selection is based on
the catalogues produced by the 
Legacy Imaging Surveys software,
\texttt{legacypipe}\footnote{\niceurl{https://github.com/legacysurvey/legacypipe}},
which uses the \texttt{Tractor} \citep{lang16a} library for source measurement.
The \texttt{legacypipe} analysis splits the sky into \emph{bricks} ($0.25^\circ \times 0.25^\circ$), and outputs products at the brick level.
The DECaLS/DR3 version was used, except for part of the NGC footprint (chunk \texttt{eboss25}), which was performed later: as the DECaLS/DR3 pipeline could not be run anymore because of a major \texttt{Python} update done on all the machines, the target selection was performed on catalogues created by the DECaLS/DR5 pipeline.
We used a slightly edited version of DECaLS/DR5 \texttt{Tractor}, using PS1 for astrometric calibration and relaxing the CCD quality cut, to prevent holes in the footprint\footnote{\niceurl{https://github.com/legacysurvey/legacypipe/tree/dr5.eboss}, \niceurl{https://github.com/legacysurvey/legacypipe/tree/dr5.eboss2}.}.
Tests on a few square degrees having the exact same exposures between DECaLS/DR3 and DECaLS/DR5 showed that $\sim$15 percent of the targets differ between the two pipeline versions.  Differences are on the faint $g$-band magnitude side of the selection, with no specific behaviour, and hence are consistent with scatter across the faint end cut.

The target selection, detailed in Table 2 of \citet{raichoor17}, consists of:
(i) a cut in the $g$-band magnitude to select \oii~emitters;
(ii) a box selection in the $grz$-diagram, with a smaller box in the NGC to prevent contamination from low-redshift objects due to shallower imaging;
(iii) a clean photometry criterion (combination of cuts on \texttt{legacypipe} output columns and of some geometrical masks).
All magnitudes are corrected for Galactic extinction with maps from \citep{schlegel98}.
We report here the magnitude cuts for SGC:
\begin{subequations}
\begin{align}
21.825 < g < 22.825\\
-0.068 \times (r-z)+0.457 < g-r <  0.112 \times (r-z)+0.773\\
0.218  \times (g-r)+0.571 < r-z < -0.555 \times (g-r)+1.901\\
\end{align}
\end{subequations}
and here the magnitude cuts for the NGC:
\begin{subequations}
\begin{align}
21.825 < g < 22.9\\
-0.068 \times (r-z)+0.457 < g-r <  0.112 \times (r-z)+0.773\\
0.637 \times (g-r)+0.399 < r-z < -0.555 \times (g-r)+1.901
\end{align}
\end{subequations}
It provides a list of 269,718 targets.

\subsection{Spectroscopic observations \label{sec:specobs}}
\begin{figure*}
\begin{center}
\includegraphics[width=0.95\textwidth]{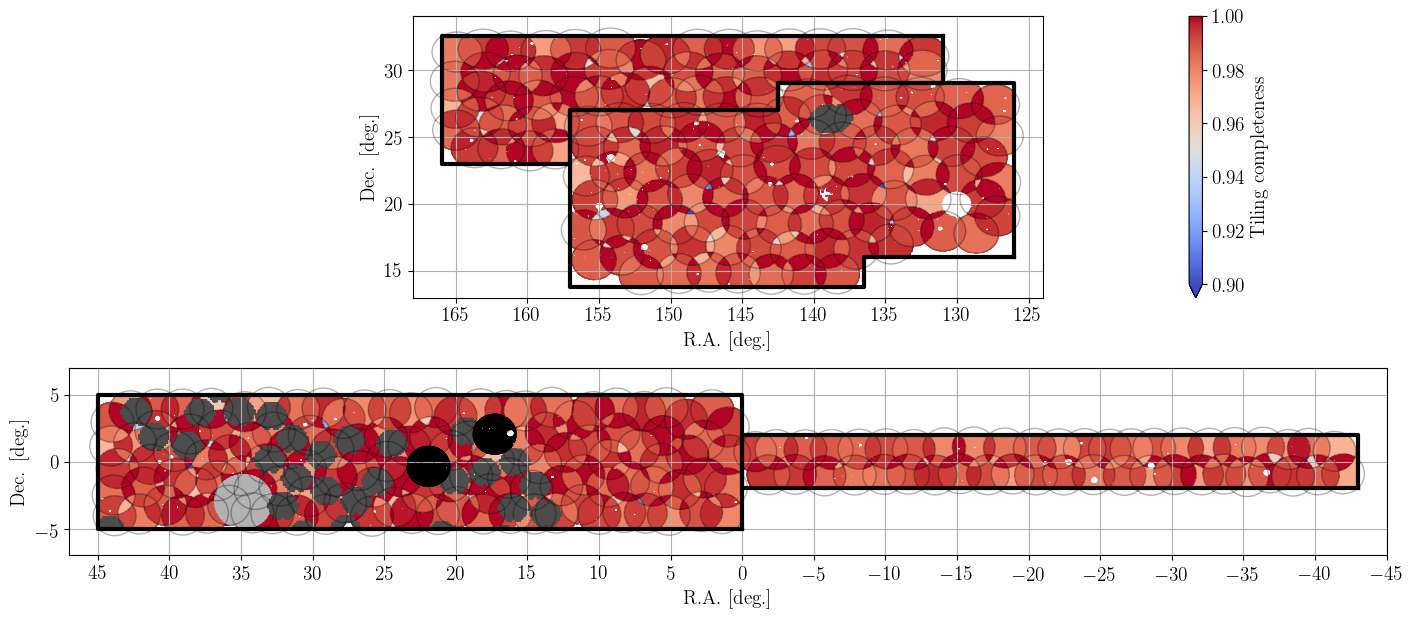}
\caption{Geometry of the ELG program. The NGC tiling is presented in the top panel: chunk \texttt{eboss23} is at lower Dec. and chunk \texttt{eboss25} at higher Dec. The SGC tiling is presented in the bottom panel: chunk \texttt{eboss21} is at R.A.<0$^\circ$. and chunk \texttt{eboss22} at R.A.>0$^\circ$.  The colour-coding is the tiling completeness (COMP\_BOSS), which represents the fraction of resolved fibres per sector (see Section \ref{sec:cp}). Additionally, we overlay some \textit{a posteriori} angular veto masks, which are detailed in Section \ref{sec:vetomask}: Mira star (light gray), DECam pointings with bad photometric calibration (dark gray), and two low-quality spectroscopic plates (black). The regions without targets at R.A$\sim$130$^\circ$ and Dec.$\sim$20$^\circ$ corresponds to the open cluster NGC 2632.}
\label{fig:tiling}
\end{center}
\end{figure*}

The ELG spectroscopic observations are conducted with the BOSS spectrograph \citep{smee13} at the 2.5-m aperture Sloan Foundation Telescope at Apache Point Observatory in New Mexico \citep{gunn06}.  1000 objects are observed at once, with 1000 fibres plugged into a drilled plate, amongst which $\sim$850 are assigned to ELGs.
305 plates have been allocated to the ELG program and observations were undertaken between Sept.~2016 and Feb.~2018 ($57656 \leq \texttt{MJD} \leq 58171$).
Targeting was performed on subsets of the full eBOSS/ELG area, called chunks: the SGC is divided in two chunks, \texttt{eboss21} and \texttt{eboss22}, and the NGC is divided in two chunks, \texttt{eboss23} and \texttt{eboss25}.
Observations are designed by defining the plate tiling \citep{blanton03b}, which optimises for each chunk the fraction of targets having a fibre for the budgeted number of plates. 
Figure \ref{fig:tiling} shows the plate tiling, with the tiling completeness, defined as the fraction of resolved targets (see Section \ref{sec:cp}, this corresponds to the COMP\_BOSS quantity in previous BOSS/eBOSS analysis).
We report in Table \ref{tab:specobs} the details of the spectroscopic observations for each chunk and for the whole programme.

Details of the spectroscopic setup are presented in \citet{raichoor17}.
Each plate is observed with individual exposures of 15 min until $\texttt{rSN}^2>22$, where $\texttt{rSN}^2$ is the median squared signal-to-noise ratio (SN) in the red camera evaluated at the mountain.
This is reached on average with 4.7$\times$15 min exposures; the average SN on individual ELG spectra is $\sim$0.8.
During the first month of operations (around half of the \texttt{eboss21} chunk), observations were done with higher $\texttt{rSN}^2$ ($\sim$40).

If one plate has to be unplugged before it reaches the minimum $\texttt{rSN}^2$, it is plugged again later and re-observed: as the fibres are not assigned to the same targets between the two pluggings, this results in two PLATE-MJD reductions for the considered plate.
This provides valuable independent, repeat observations for ELGs on that plate, which allows us to quantify the reliability of our redshift measurement (see Section \ref{sec:redrock}).

Because of dead fibres or observational issues (e.g., incorrect plugging of a fibre), some spectra are unusable.
We identify those cases by using the \texttt{ZWARNING} quantity output by the redshift fitter \citep[see Table 3 of ][]{bolton12}: when one of the \texttt{LITTLE\_COVERAGE}, \texttt{UNPLUGGED}, \texttt{BAD\_TARGET}, or \texttt{NODATA} bits is turned on, we label the fibre as not valid, and as a consequence we discard the spectrum and consider that no spectroscopic observation has been taken.

Overall, there are 14,799 repeat ELG spectra, or duplicates.
Duplicates happen for two reasons.
First, when a PLATE has several MJD reductions: all ELGs on the plate will have as many $\zspec$ measurements as MJD reductions.
In that case, we consider as primary spectra all spectra coming from the MJD reduction with the higher plate SN, and as duplicates the spectra from the other MJD reductions.
Second, in the plate overlap regions, any remaining fibres are assigned to repeats: the fibre is then assigned to a target which already has a fibre assigned from another overlapping plate.
In that case, we consider as primary the spectrum with a valid fibre and with the highest $\chi^2$ difference between the best-fit solution and the second-best fit solution.

\begin{table*}
\centering
\caption{Spectroscopic observations properties per chunk: (1): chunk name; (2): tiling area [deg$^2$]; (3): number of plates; (4): number of PLATE-MJD reductions; (5): average observed time in minutes per PLATE-MJD; (6): average observed time in minutes included in the reduction per PLATE-MJD; (7): mean plate rSN$^2$; (8): mean SN per spectrum; (9): number of targets; (10): number of observed spectra; (11): number of spectra after removing duplicates; (12): number of targets after applying the veto LSS masks; (13): number of star spectra after applying the veto LSS masks; (14): number of galaxy spectra after applying the veto LSS masks; (15): number of galaxy spectra after applying the veto LSS masks and with a reliable redshift.}
\begin{tabular}{lcccccccccccccc}
\hline
\hline
(1) & (2) & (3) & (4) & (5) & (6) & (7) & (8) & (9) & (10) & (11) & (12) & (13) & (14) & (15)\\
Chunk & Area & $N_{\rm PLATE}$ & $N_{\rm PLATE}^{\rm MJD}$ & $t_{\rm exp}^{\rm obs}$ & $t_{\rm exp}^{\rm kept}$ & rSN$^2$ & SN$_{\rm spec}$ & $N_{\rm targ}$ & $N_{\rm spec}^{\rm obs}$ & $N_{\rm spec}^{\rm obs,uniq}$ & $N_{\rm targ}^{\rm LSS}$ & $N_{\rm spec}^{\rm LSS,star}$ & $N_{\rm spec}^{\rm LSS,gal}$ & $N_{\rm spec,reliable}^{\rm LSS,gal}$\\
 & [deg$^2$] &  &  & [min] & [min] &  &  &  &  &  & &  &  & \\
\hline
\texttt{eboss21} & 171 & 46 & 46 & 122 & 100 & 28.7 & 0.99 & 40904 & 38992 & 38493 & 36314 & 333 & 33884 & 31200\\
\texttt{eboss22} & 445 & 121 & 131 & 86 & 73 & 22.1 & 0.85 & 106897 & 111061 & 101954 & 79880 & 512 & 75585 & 69071\\
\texttt{eboss23} & 377 & 87 & 92 & 70 & 60 & 25.4 & 0.82 & 76236 & 76250 & 71134 & 70935 & 544 & 65677 & 58648\\
\texttt{eboss25} & 178 & 51 & 51 & 59 & 54 & 24.6 & 0.81 & 45141 & 42940 & 42863 & 42565 & 315 & 40141 & 36166\\
\texttt{all} & 1170 & 305 & 320 & 82 & 70 & 24.0 & 0.84 & 269178 & 269243 & 254444 & 229694 & 1704 & 215287 & 195085\\
\hline
\end{tabular}
\label{tab:specobs}
\end{table*}

\subsection{Spectroscopic redshift estimation: \texttt{redrock} \label{sec:redrock}}

The results presented in this paper use version v5\_13\_0 of the \texttt{idlspec2d} data reduction pipeline to extract and flux-calibrate the ELG 1D spectra from the raw 2D spectroscopic images \citep{bolton12,ahumada19}.
As stated in \citet{raichoor17}, the BOSS/eBOSS redshift fitter, \texttt{idlspec1d}, is not optimised for ELGs, as it has been designed for bright LRGs.
Therefore, we used for the 1D spectrum analysis \texttt{redrock}\footnote{\niceurl{https://github.com/desihub/redrock}; we used a customed version of the tagged version 0.14.0, where we do not use the ANDMASK masking, as it unnecessarily removes pixels close to sky emission lines from the fit, hence creating artificial drops in the redshift density n($z$), where the \oii~doublet falls close to sky lines; that version is internally labelled v5\_13\_0\_no\_andmask.}, the DESI redshift fitter, which provides more reliable redshifts.

We present here a summary of the \texttt{redrock} principle; we refer the reader to \citet{ross20} for more details.
\texttt{redrock} templates, labelled archetypes, are the most representative (simulated) physical spectra of DESI galaxies, QSOs, and stars.
\texttt{redrock} fitting procedure includes two steps.
In the first step, it finds the $\chi^2$ minima using PCA templates, based on DESI archetypes.
As the best-fit PCA spectra can be non-physical, for each minimum vicinity, \texttt{redrock} then recomputes the $\chi^2$ with archetypes.
This approach ensures that the best-fit solution corresponds to a physical, meaningful, spectrum.

Following the eBOSS requirements \citep{dawson16,raichoor17}, redshift estimates should be precise ($|\Delta v|<300$ \kms) and accurate (less than 1 percent catastrophic redshifts, defined as $|\Delta v|>1000$ \kms).
To match these requirements, we define a redshift estimate reliable if the following criteria are satisfied:
\begin{subequations}
\begin{align}
(\texttt{ZWARNING}==0) \; \textrm{and} \label{eq:zreliable_zwarn} \\
(\texttt{SN\_MEDIAN}[i]>0.5 \; \textrm{or} \; \texttt{SN\_MEDIAN}[z]>0.5) \; \textrm{and} \label{eq:zreliable_sn} \\
(\texttt{zQ}>=1 \; \textrm{or} \; \texttt{zCont}>=2.5)  \label{eq:zreliable_zQCont} 
\end{align}
\end{subequations}
The first criterion (\ref{eq:zreliable_zwarn}) is based on the \texttt{ZWARNING} flag output by \texttt{redrock} (see Section \ref{sec:specobs}) and ensures that the fitting did not encounter any problems.
In particular, it assures that the coefficient in front of the best-fitting archetype spectrum is positive, meaning that the best-fit template is physically motivated \citep[see][]{ross20}.
The second criterion (\ref{eq:zreliable_sn}) ensures a minimum SN in the red part of the spectrum, where the \oii~ line is expected to be observed at $z \sim 1$\footnote{\texttt{SN\_MEDIAN}[i,z] is the median SN for all good pixels from the spectrum corresponding to the $i$- and $z$-band.}.
The third criterion (\ref{eq:zreliable_zQCont}) reduces the fraction of catastrophic redshifts; it is based on the  \{\texttt{zQ, zCont}\} \textit{a posteriori} flags \citep[see][]{comparat16a,raichoor17}, which quantify the emission lines and continuum level of information.
The impact of each cut, along with the improvement with respect to \texttt{idlspec1d} are shown in \mbox{Table \ref{tab:zreliable}} (the catastrophic rate is estimated with repeat observations, as described further in this Section).
One can see the significant improvement brought by \texttt{redrock} with respect to the reliability criterion presented in \citet{raichoor17}, based on \texttt{idlspec1d}: it allows us to include in our cosmological $0.6<\zspec<1.1$ sample more reliable redshifts (80.7 percent vs. 74.0 percent, for a Poissonian fluctuation of $\sim$0.3 percent), with a lower fraction of catastrophic rate (0.3 percent vs. 0.5 percent, for a Poissonian fluctuation of $\sim$0.06 percent).
Those improvements are significant, well above the Poissonian noise fluctuations.
We validate our reliability criteria with two approaches, visual inspection and repeat observations.

\begin{table*}
\setlength{\tabcolsep}{3pt}
\centering
\caption{Reliable redshift statistics for various criteria. We use the last line criterion. Estimate from our catastrophic rates are computed from repeat observations; see Table \ref{tab:zspecvi} for our visual inspection results.}
\begin{tabular}{llcccc}
\hline
Redshift & criterion & reliable & reliable & catastrophic & catastrophic \\
fitter & &  $\zspec$  &  $0.6<\zspec<1.1$ & $\zspec$ & $0.6<\zspec<1.1$\\
\hline
\texttt{idlspec1d} & Eq.~(1) of \citet{raichoor17} & 83.1\% & 74.0\% & 0.5\% & 0.5\% \\
\texttt{redrock} & Eq.~(\ref{eq:zreliable_zwarn}) & 93.0\% & 82.0\% & 0.7\% & 0.6\% \\
\texttt{redrock} & Eq.~(\ref{eq:zreliable_zwarn}) \& Eq.~(\ref{eq:zreliable_sn})& 91.8\% & 81.3\% & 0.6\% & 0.6\% \\
\texttt{redrock} & Eq.~(\ref{eq:zreliable_zwarn}) \& Eq.~(\ref{eq:zreliable_sn}) \& Eq.~(\ref{eq:zreliable_zQCont}) & 90.6\% & 80.7\%  & 0.3\% & 0.3\% \\
\hline
\end{tabular}
\label{tab:zreliable}
\end{table*}

\begin{table}
\centering
\caption{Redshift measurement assessment from visual inspection of three plates for $\sim$1900 ELGs, with $0.6<\zspec<1.1$ and passing Eqs.~(\ref{eq:zreliable_zwarn}), (\ref{eq:zreliable_sn}), (\ref{eq:zreliable_zQCont}). The visual inspection confidence flag meaning is:
3: definitely correct, 
2: features are visible and the redshift is likely to be correct, 
1: information in the spectrum, but the redshift is a guess,
0: no information, useless spectrum.
For instance, 24.0 percent of the inspected spectra have confidence=2, and 99.3 percent of those have $|\Delta v|<300$ \kms.}
\begin{tabular}{lccc}
\hline
Conf. Flag & Percentage & $|\Delta v|<300$ \kms & $|\Delta v|<1000$ \kms \\
\hline
3 & 71.5\%  & 99.9\% & 99.9\%\\
2 & 24.0\%  & 99.3\% & 99.6\%\\
1 & 2.9\%   & 94.5\% & 96.3\%\\
0 & 1.6\%   &  6.5\%&  6.5\%\\
all & 100\% & 98.1\% & 98.2\%\\
\hline
\end{tabular}
\label{tab:zspecvi}
\end{table}

Three plates have been visually inspected, one from the eBOSS/ELG program (PLATE-MJD=9236-57685) and two from pilot ELG programs (PLATE-MJD=6931-56388 and 8123-56931).
We restrict here to the $\sim$1900 ELGs with $0.6<\zspec<1.1$ that passed our reliable criteria listed in  Eqs.~(\ref{eq:zreliable_zwarn}), (\ref{eq:zreliable_sn}), and (\ref{eq:zreliable_zQCont}).
The inspector assigns a visual redshift and one of the following confidence flags:
3: definitely correct, 
2: features are visible and the redshift is likely to be correct, 
1: information in the spectrum, but the redshift is a guess,
0: no information, useless spectrum.
Visual inspection results are reported in Table \ref{tab:zspecvi}.
The \texttt{redrock} redshift is almost in perfect agreement with the inspector redshift for confidence=3 and confidence=2 (95.5 percent of the sample).
For confidence=1 (2.9 percent of the sample), both redshift estimations mostly agree ($\sim$95 percent).
For confidence=0 (1.6 percent of the sample), we can conservatively assume that the pipeline is wrong in most cases.
Overall, based on these visual inspections we estimate that the pipeline provides a redshift precision better than 300 \kms for 98.1 percent of our sample and a  catastrophic redshift for $\sim$1.8 percent of our sample.

We present a second, independent estimate of catastrophic rate with repeat observations, which provides us with $\sim$17,000 pairs of observations of a given target.
We restrict to the $\sim$13,000 repeats where both redshift estimations pass our reliability criterion, and consider a pair is catastrophic if the two redshift measurements differ by more than 1000 \kms.
Following this approach, we find that 0.3 percent of the sample have a catastrophic redshift measurement.
Additionally, we can assess with repeats that 99.5, 95, and 50 percent of our redshift estimates have a precision better than 300 \kms, 100 \kms, and 20 \kms, respectively.

We thus conclude that the \texttt{redrock} redshift measurement is reliable, with a precision better than 300 \kms for $\sim$99 percent of our sample and an expected catastrophic rate of $\sim$1 percent, thus fulfilling the eBOSS/ELG requirements set at the beginning of the program.

\section{Large-scale structure catalogues creation} \label{sec:lss}

We detail in this section the building of the LSS catalogues.
These LSS catalogues are used in this paper to measure the spherically averaged BAO in configuration space.
They are also used in \citet{de-mattia20} and \citet{tamone20} for the measurement of the growth rate of structures and BAO in Fourier space and in configuration space, respectively.
They are publicly available\footnote{A link to webpage will be provided after DR16 papers are accepted for publication.}.

Table \ref{tab:lssprop} summarises the overall properties of these LSS catalogues.
The steps to build the LSS catalogues are:
1) define starting data and random samples; 
2) define and apply the angular veto masks to the data and the randoms; 
3) define weights to correct for non-cosmological fluctuations (redshift failures: $w_{\rm noz}$, close pairs: $w_{\rm cp}$, systematics due to photometry: $w_{\rm sys}$), and to optimise the contribution of galaxies based on their number density at different redshifts and apply inverse-variance weights $w_{\rm FKP}$;
4) assign redshifts to the randoms.

\begin{table}
\centering
\caption{Statistic for the ELG sample. The reported $N$ are computed after applying the LSS veto masks. A target is either observed or unobserved because of close pairs or lack of fibre: $N_{\rm obs} + N_{\rm cp} + N_{\rm miss} = N_{\rm targ}$. Similarly, an observed target is classified as a star, as a galaxy, or a redshift failure: $N_{\rm star} + N_{\rm gal} + N_{\rm zfail} = N_{\rm obs}$. $N_{\rm used}$ is the number of galaxies with $0.6<\zspec<1.1$.
The geometric area is the tiling area, i.e. covered by the plates.
The unvetoed area is the area after applying the LSS veto masks.
The effective area is the unvetoed area after accounting for the tiling  completeness.}
\begin{tabular}{lrrr}
\hline
\hline
 & NGC & SGC & Total\\
\hline
$N_{\rm targ}$  &   113,500 &   116,194     &   229,694\\
$N_{\rm obs}$   &   106,677 &   110,314     &   216,991\\
$N_{\rm cp}$    &   5,805   &   4,797       &   10,602\\
$N_{\rm miss}$  &   1,018   &   1,083       &   2,101\\
$N_{\rm gal}$   &   94,814  &   100,271     &   195,085\\
$N_{\rm star}$  &   859     &   845         &   1,704\\
$N_{\rm zfail}$ &   11,004  &   9,198       &   20,202\\
$N_{\rm used}$  &   83,769  &   89,967      &   173,736\\
Geometric area [deg$^2$]    & 554.1  &   616.1  &   1170.2\\
Unvetoed area  [deg$^2$]    & 372.8  &   360.9  &   733.8\\
Effective area  [deg$^2$]   & 369.5  &   357.5  &   727.0\\
\hline
\end{tabular}
\label{tab:lssprop}
\end{table}

\subsection{Data selection, random catalogues \label{sec:datarand}}
To construct the LSS catalogues, we first remove duplicates and restrict to ELGs with a valid fibre and a reliable $\zspec$ estimate with $0.6<\zspec<1.1$: this provides 173,736 unique ELGs.

We generate random catalogues (randoms), which will have the same angular and radial distribution as the ELG data.
We first create random angular positions at a constant angular density of 10$^4$ deg$^{-2}$, i.e. $\sim$40$\times$ the ELG target density, over the full sky.  We then remove any random outside of any chunk.

\subsection{Angular veto masks \label{sec:vetomask}}
In addition to the geometry of the plate tiling, we apply several angular veto masks to our LSS data and random catalogues where, for various reasons, we could not reliably observe galaxies.
Table \ref{tab:vetomask} lists all those angular veto masks, along with the masked area and the number of masked targets.

\begin{table}
\centering
\caption{Angular veto mask properties. bits 1,2,3,4 and 5 have been applied before the target selection (the few removed targets are due to slightly different implementation). Apart from the \texttt{eboss22} two low-quality plates removal, all veto masks are bit-coded (in the mskbit column in the catalogues).}
\begin{tabular}{llcc}
\hline
\hline
bit & mask & removed area & removed targets\\
 & & [deg$^2$] & \\
\hline
1 & not g+r+z & 67.2 & 27\\
2 & xybug & 49.7 & 0\\
3 & recovered \texttt{decam\_anymask} & 210.1 & 142\\
4 & \texttt{tycho2inblob} & 4.7 & 0\\
5 & bright objects & 57.6 & 7\\
6 & \textit{Gaia} stars & 54.0 & 17456\\
7 & Mira star & 12.5 & 3555\\
8 & imprecise mskbit 3 & 0.1 & 15\\
9 & centerpost & 0.6 & 166\\
10 & TDSS\_FES targets & 1.3 & 308\\
11 & DECam bad phot. calib. & 72.7 & 16325\\
- & \texttt{eboss22} low-quality plates & 13.9 & 3123\\
\hline
- & total & 436.5 & 41124\\
\hline
\end{tabular}
\label{tab:vetomask}
\end{table}

Masks corresponding to bit values 1 to 5 in Table \ref{tab:vetomask} were applied at the target selection level, before the tiling \citep[those are described in][]{raichoor17}.
The other masks are applied in the analysis step, after the spectroscopic observations: those additional angular masks remove a significant number of targets, but are necessary to provide a clean, reliable LSS catalogue.

Masks corresponding to bit values 1 through 4 rely on the photometric \texttt{legacypipe} pipeline outputs, stored (or recovered for bit=3) at the brick level.
Those outputs are the photometric catalogues, but also various brick-sized images ($3600 \times 3600$ pixels, with 0.262 arcsec/pixel), such as the depth images.
We detail below each veto angular mask.

\begin{itemize}

\item not g+r+z (bit=1):
the target selection requires that $grz$-photometry is available: this \textit{de facto} excludes regions not covered by $grz$-imaging.
Those regions can be identified with the \texttt{legacypipe} depth images;

\item(x,y) bug (bit=2):
as stated in \citet{raichoor17}, a bug at the target selection level resulted in an additional angular masking.
This affects the \texttt{eboss23} chunk, but also -- to a lesser extent -- the \texttt{eboss21} and \texttt{eboss22} chunks; the \texttt{eboss25} chunk is not affected by this mask.
This mask can be exactly recovered with using the \texttt{legacypipe} depth images;

\item \texttt{decam\_anymask} (bit=3):
in the target selection, we required $\texttt{decam\_anymask[grz]}=0$, where \texttt{decam\_anymask} is a \texttt{legacypipe} quantity, flagging objects where one of the underlying DECam images is defective at the pixel position corresponding to the center of the object; this flag is often turned on for pixels close to individual imaging CCD edges along the R.A..
In the DECaLS/DR3 version, the \texttt{decam\_anymask} information is stored only where objects are detected, making it extremely difficult to propagate that information to the random sample; however, since the DECaLS/DR7 version, this information is stored at the pixel level for each brick, making it recoverable at any location.
We thus re-run the part of the DECaLS/DR7 pipeline on the exact DECam imaging dataset used for the ELG target selection (smaller than the DECaLS/DR7 one) to produce that output, having in this way the \texttt{decam\_anymask} information at the pixel level;

\item \texttt{tycho2inblob} (bit=4):
in the target selection, we required $\texttt{tycho2inblob}=\texttt{False}$,
where \texttt{tycho2inblob} is a \texttt{legacypipe} column flagging objects whose light profile overlaps one of the Tycho2 stars \citep{hog00}.
The \texttt{legacypipe} pipeline stores for each brick that information;

\item bright objects and Tycho2 stars (bit=5):
we used geometrical masks to veto the surrounding area of SDSS bright objects\footnote{\niceurl{https://data.sdss.org/sas/dr10/boss/lss/reject\_mask/}}; we also define a circular mask for each 0 mag $< V < $11.5 mag Tycho2 star with radius = $10^{3.5-0.15 \times V}$ arcsec, where $V$ is the Tycho2 star \texttt{MAG\_VT} quantity from \citet{hog00};

\item \textit{Gaia} stars (bit=6):
The \textit{Gaia}/DR2 release \citep{gaia-collaboration18} allows one to select a clean star sample for $12 < G <17$, where it is complete\footnote{\niceurl{https://www.cosmos.esa.int/web/gaia/dr2}}, hence nicely completing the Tycho2 star sample.
After defining a criterion to identify stars\footnote{If we note gmag and excess the \texttt{PHOT\_G\_MEAN\_MAG} and \texttt{astrometric\_excess\_noise} quantities, our criterion is: excess=0 or log$_{10}$(excess)<0.3 $\cdot$ gmag-5.3 or log$_{10}$(excess)<-0.5 $\cdot$ gmag+9.0.}, we group the selected stars in 1 magnitude bins, and, for each bin, analyse the ELG target density and the SSR (Spectroscopic Success Rate defined in Eq.~\ref{eq:ssr}) as a function of the distance to the stars.
We observe that, close to \textit{Gaia} stars, we select more targets, have more failures, and the redshift distribution is different: it is very likely that the excess targets correspond to artefacts in the DECaLS imaging or real objects with unreliable photometry, hence increasing the target density and the failure rate, and changing the redshift distribution.
We define a circular mask for each \textit{Gaia} star with $0<G<16$ with radius = $10^{2.32-0.07 \times G}$ arcsec, chosen by analysing the variations of the target density, the redshift failure rate, and the redshift distribution.

\item Mira star (bit=7):
The Mira star (R.A=34.84$^\circ$,Dec.=$-2.98^\circ$) is a well-known variable star, with a variability amplitude of several magnitudes.
As a consequence, its magnitude in the Tycho2 catalogue is not representative of its magnitude during the DECam observations.
We conservatively use a circular mask with a 2 degree radius around the Mira star.
This mask is displayed in light gray in Figure \ref{fig:tiling};

\item imprecise recovered \texttt{decam\_anymask} (bit=8):
our approach to recover the \texttt{decam\_anymask} value at each position of the sky to apply the bit=3 masking does not perfectly match the DECaLS catalogues used for target selection, i.e. it does not perfectly reproduces what has been used at the target selection level.
We account for this issue as follows.
We use the \texttt{Healpix}\footnote{\niceurl{http://healpix.jpl.nasa.gov}} scheme \citep{gorski05} to divide the sky into equal-area small pixels of $\sim$11 arcmin$^2$ (corresponding to nside = 1024).
We reject 37 pixels where the percentage of objects with an improper recovered \texttt{decam\_anymask} is greater than 10 percent;

\item centerpost (bit=9):
each plate has a hole in its centre to fix it with the centrepost; as a consequence, no fibre can be placed within 92\arcsec~ of the plate centre.
Contrary to other BOSS/eBOSS targets, the higher ELG density making the tiling denser, this does not result in a ``simple'' veto mask, as the position of plate centre can be covered by another adjacent plate (see Figure \ref{fig:tiling}).
However, for simplicity, we simply mask these centerpost regions;

\item TDSS FES targets (bit=10):
on each ELG plate, $\sim$50 fibres are assigned to the Time Domain Spectroscopic Survey \citep[TDSS,][]{morganson15,ruan16}.
A subsample of the TDSS targets, the FES class targets ($\sim$1 deg$^{-2}$), have been tiled with the same priority as the ELG targets.
To account for that, we create around each TDSS FES target a circular veto mask with a radius of 62 arcsec, corresponding to the size of one fibre;

\item DECam bad photometric calibration (bit=11):
at the time of DECaLS/DR3, the DECaLS pipeline was including all public $grz$-band DECam imaging over the DECaLS footprint, hence imaging from various different programs.
The latest DECaLS/DR8 release\footnote{\niceurl{http://legacysurvey.org/dr8}} mostly restricts to DES and DECaLS observations, and has a significantly improved photometric calibration procedure.
We take advantage of that dataset to verify the photometric calibration of our DECaLS/DR3 and DR5 imaging used for target selection.
We identify in this way some observing programs with improper photometric calibration (of the order of tens of mmag): such systematic offsets in the photometry implies a different target selection, as it is equivalent to move the boundaries of the photometric cuts.
We remove the regions covered by the DECam CCDs belonging to those identified observing programs.
This mask is displayed in dark gray in Figure \ref{fig:tiling};

\item \texttt{eboss22} low-quality plates:
lastly, we also remove the regions covered by two \texttt{eboss22} spectroscopic plates (PLATE-MJD=9430-58112 and 9395-58113), which have significantly lower-than-average quality.
Those plates bias the SSR=f(pSN) fit in Eq.~\ref{eq:zfailpsn} (see next Section).
This mask is displayed in black in Figure~\ref{fig:tiling}.

\end{itemize}

\begin{figure}
\begin{center}
\includegraphics[width=0.95\columnwidth]{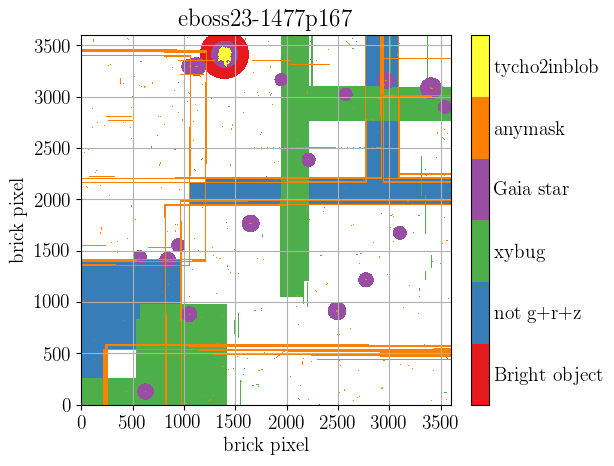}
\caption{Illustration of the DECaLS-related, bright objects and stars masks for a given DECaLS brick ($0.25^\circ \times 0.25^\circ$, $3600 \times 3600$ pixels, with 0.262 arcsec/pixel).
The xybug mask is the symmetric along the brick diagonal of the not g+r+z mask.
The \texttt{decam\_anymask} mask mostly follows the CCD edges along R.A. (horizontal in the figure).}
\label{fig:maskbrick}
\end{center}
\end{figure}

Figure \ref{fig:maskbrick} illustrates the DECaLS-related, bright objects and stars masks for a given DECaLS brick.

We provide in the associated data release the required information to reproduce the angular masking when considering any (R.A., Dec.) position: bits 1 to 7 can computed with the \texttt{brickmask}\footnote{\niceurl{https://github.com/cheng-zhao/brickmask/releases/tag/v1.0}} script, bits 8 to 11 and the two \texttt{eboss22} low-quality plates can be reproduced with customed python lines.

\subsection{Spectroscopic redshift failures}
The principle of using ELGs for large-scale structure clustering relies on the fact that it is possible to measure the $\zspec$ thanks to emission lines, with no requirement of high SN detection of the continuum, making them an interesting tracer.
However, for low SN spectra (see Table \ref{tab:specobs}), the BOSS spectrograph resolution of $\sim$2000 does not allow the \oii~doublet to be resolved \citep{comparat13,comparat13a}, on which many $\zspec$ measurements rely.
As a consequence, redshift failures are significant ($\sim$10 percent of the observations), and present strong dependencies on observing conditions, which need to be carefully modelled and corrected for in the large-scale structure analysis \citep[see also][]{bautista18}.

We define the Spectroscopic Success Rate (SSR) as:
\begin{equation}
\text{SSR} = \frac{N_{\text{gal}}}{N_{\text{gal}} + N_{\text{zfail}}},
\label{eq:ssr}
\end{equation}
where $N_{\rm gal}$ is the number of spectroscopic spectra with a valid fibre, a reliable $\zspec$ estimate, and not being a star, and $N_{\rm zfail}$ is the number of spectra with a valid fibre but no reliable $\zspec$ estimate and not a star.
We beforehand apply all angular veto masks described in Section \ref{sec:vetomask}.

To correct redshift failures, we derive weights from a fit of the SSR as a function of two quantities, which correlate with the angular position of the fibres on the sky, namely the plate-average SN (pSN) and the (XFOCAL, YFOCAL) position in the focal plane:
\begin{equation}
w_{\text{noz}} = \frac{1}{f_{\text{noz,pSN}} \cdot f_{\text{noz,XYFOCAL}}},
\label{eq:weightnoz}
\end{equation}

\begin{figure}
\begin{center}
\includegraphics[width=0.95\columnwidth]{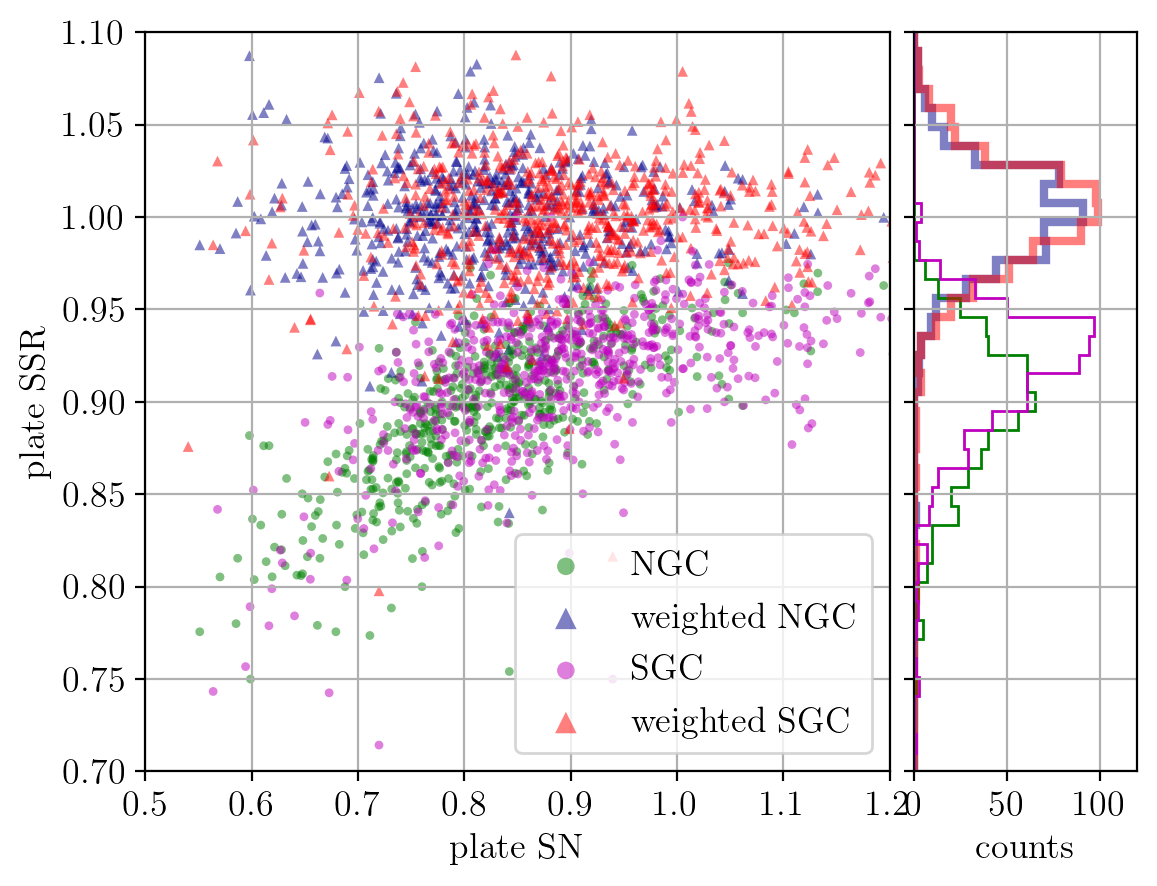}
\caption{Fraction of reliable $\zspec$ (SSR) per plate, as a function of the plate SN: each dot represent a PLATE-MJD reduction. For the NGC/SGC, the SSR before weighting by $1/f_{\text{noz,pSN}}$ is displayed in green/magenta dots and the SSR after weighting by $1/f_{\text{noz,pSN}}$ is displayed in blue/red triangles. The model is fitted to each half-spectrograph for each chunk.}
\label{fig:zfailpsn}
\end{center}
\end{figure}

\begin{figure}
\begin{center}
\begin{tabular}{c}
\includegraphics[width=0.95\columnwidth]{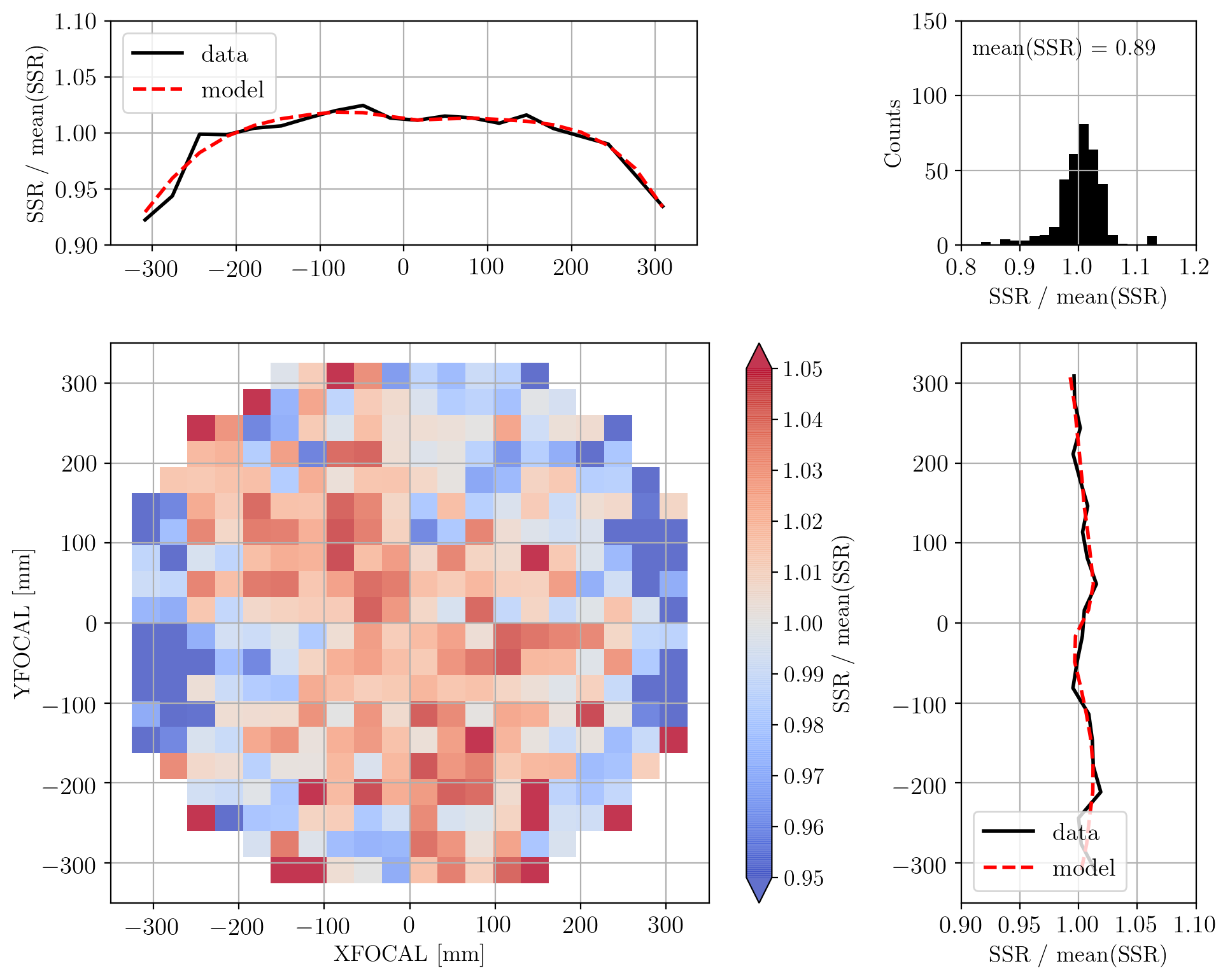}\\ 
\\[20pt]
\includegraphics[width=0.95\columnwidth]{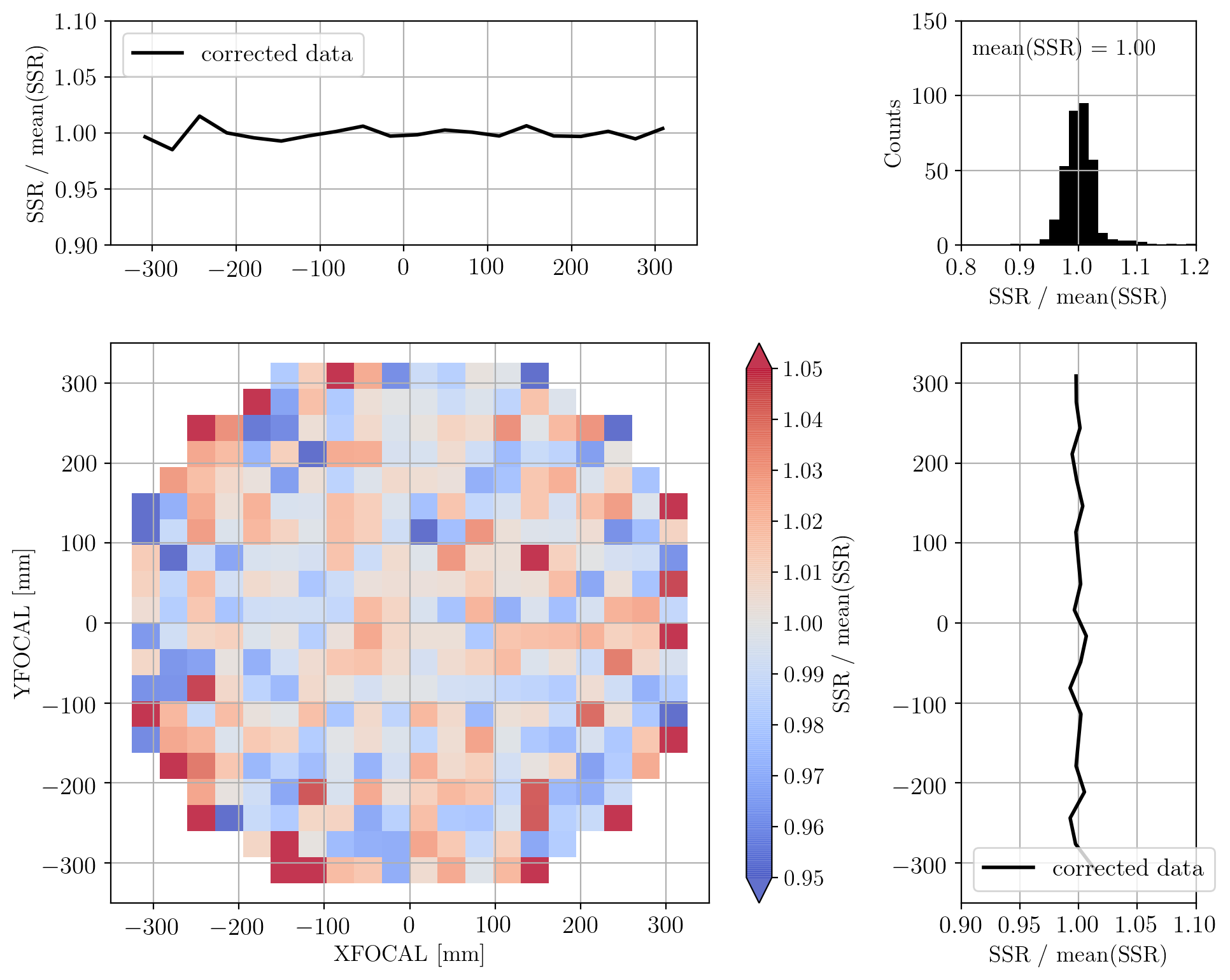}\\
\end{tabular}
\caption{Fraction of reliable $\zspec$ (SSR)  as a function of XFOCAL and YFOCAL for the NGC, before (top panels) and after (bottom panels) weighting by $1/f_{\text{noz,XYFOCAL}}$. The top- and right-side panels show the SSR as a function of XFOCAL and YFOCAL; the top-right histograms display the distribution of the normalised SSR. The model is fitted to each half-spectrograph for each chunk.}
\label{fig:xyfoc_ngc}
\end{center}
\end{figure}

\begin{figure}
\begin{center}
\begin{tabular}{c}
\includegraphics[width=0.95\columnwidth]{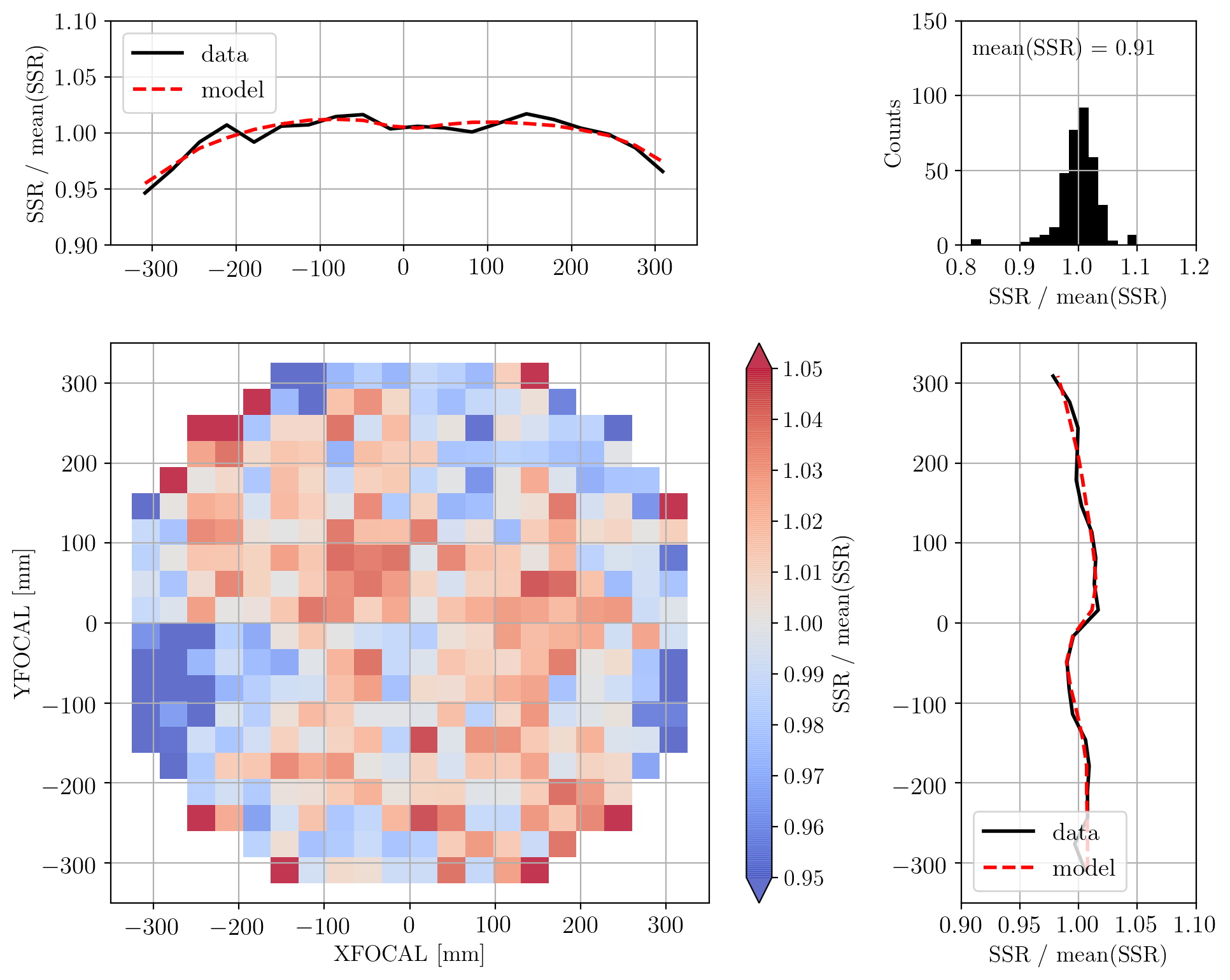}\\
\\[20pt]
\includegraphics[width=0.95\columnwidth]{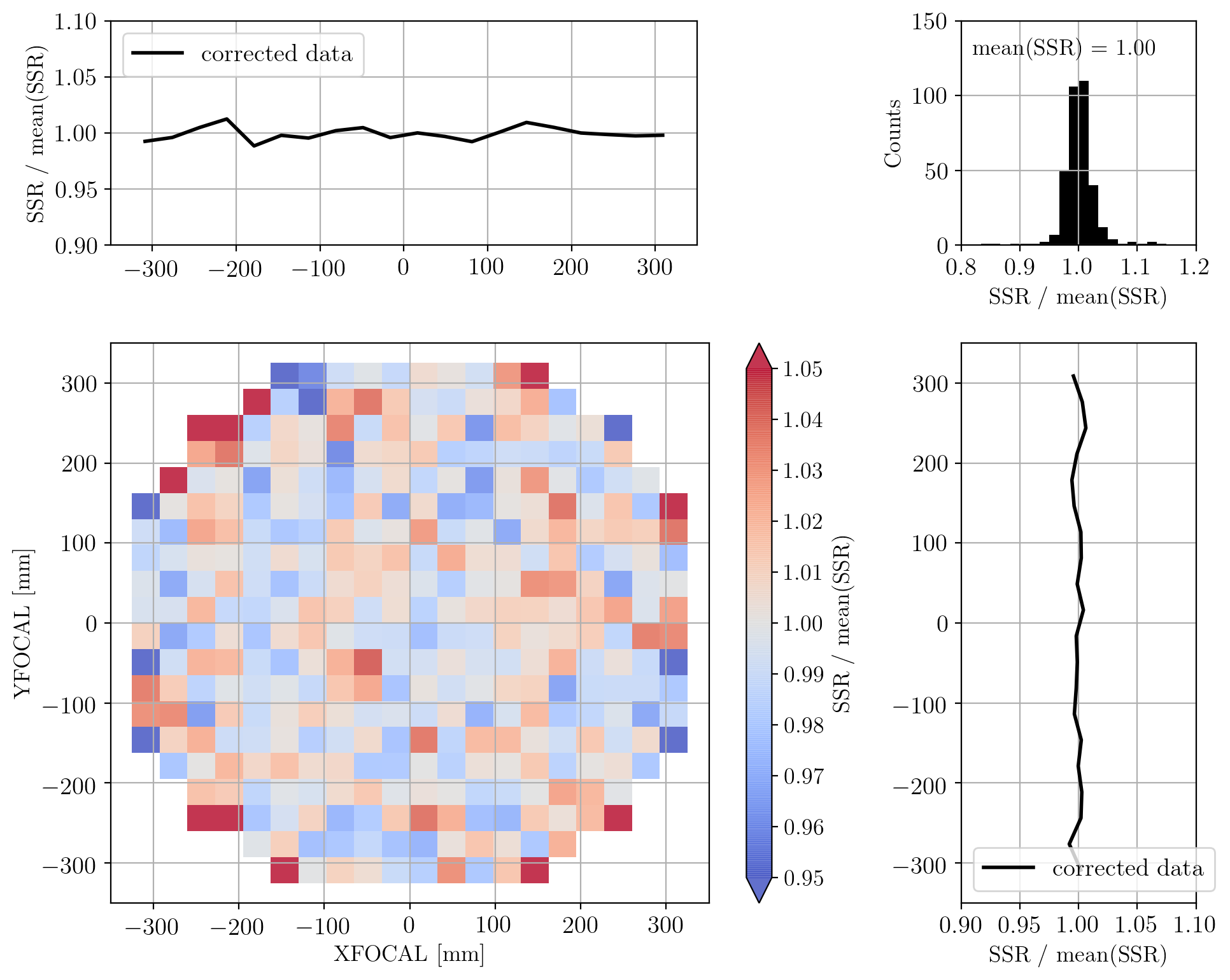}\\
\end{tabular}
\caption{Same as Figure \ref{fig:xyfoc_ngc}, but for the SGC.}
\label{fig:xyfoc_sgc}
\end{center}
\end{figure}

We perform the fit for each half-spectrograph (Spectro\_1a: $1 \leq \texttt{FIBERID} \leq 250$, Spectro\_1b: $251 \leq \texttt{FIBERID} \leq 500$, Spectro\_2a: $501 \leq \texttt{FIBERID} \leq 750$, Spectro\_2b: $751 \leq \texttt{FIBERID} \leq 1000$) of each chunk (\texttt{eboss21},\texttt{eboss22}, \texttt{eboss23}, \texttt{eboss25}).
The rationale behind this approach stems from the specificity of each chunk and the different response of each half-spectrograph.
Indeed, \texttt{eboss21} has longer spectroscopic exposure time on average and a particular geometry (hence having a non-standard position of the fibres in the focal plane), \texttt{eboss21} and \texttt{eboss22} have DES, deeper imaging, while \texttt{eboss23} imaging is shallower and \texttt{eboss25} imaging comes from a different DECaLS release.
It is known that the second spectrograph ($501 \leq \texttt{FIBERID} \leq 1000$) has a better throughput \citep{smee13}: we do observe differences due to this for our ELG sample, and we also observe that half-spectrographs have different responses; for instance the mean SN per spectra is 0.91, 0.87, 0.94, 0.88 for Spectro\_1a, Spectro\_1b, Spectro\_2a, Spectro\_2b, respectively.
We currently do not find an explanation for that half-spectrograph difference in the mean SN.
For simplicity, we display in Figures \ref{fig:zfailpsn}, \ref{fig:xyfoc_ngc} and \ref{fig:xyfoc_sgc} the fitted results for all fibres from each Galactic cap.

The first quantity is the overall SN of the plate, pSN.
As observations are performed at a rather low SN, the fraction of redshift failures increases quickly for lower-than-average observing conditions.
In Figure \ref{fig:zfailpsn} we display the plate SSR, i.e. the fraction of reliable $\zspec$ per plate, as a function of the plate SN, being defined as the average ELG SN on the plate.
We model the SSR dependence on the plate SN with the following function:
\begin{equation}
f_{\text{noz,pSN}}(x) = c_0 - c_1 \times |x-c_2|^{c_3},
\label{eq:zfailpsn}
\end{equation}
where $x$ is the pSN and the four coefficients $c_0, c_1, c_2$, and $c_3$ are fitted through a $\chi^2$ minimisation.
For each fit, the number of fitted points is the number of plates per chunk, reported in column (3) of Table \ref{tab:specobs}.
Figure \ref{fig:zfailpsn} illustrates how the data populate the {pSN, SSR} space, before (dots) and after (triangles) the weighting by $1/f_{\text{noz,pSN}}$.
Once weighted, the SSR is independent of the plate SN.

The second quantity we use is the (XFOCAL,YFOCAL) position.
On average, fibres from Spectro\_1a are at YFOCAL<0, XFOCAL>0, from Spectro\_1b at YFOCAL<0, XFOCAL<0, from Spectro\_2a at YFOCAL>0, XFOCAL<0, and from Spectro\_2b at YFOCAL>0, XFOCAL>0.
We model the SSR dependence on (XFOCAL,YFOCAL) with the following function:
\begin{equation}
f_{\text{noz,XYFOCAL}}(x,y) = c_0 - c_1 \times |x-c_2|^{c_3} - c_4 \times |y-c_5|^{c6},
\label{eq:zfailxyfoc}
\end{equation}
where ($x,y$) are the centre coordinates of bins in the (XFOCAL,YFOCAL) plane, and the seven coefficients $c_0, c_1, c_2, c_3, c_4, c_5$, and $c_6$  are fitted through a $\chi^2$ minimisation.
For each fit, the number of fitted points is $\sim$350, the number of bins in the (XFOCAL,YFOCAL) plane.
Figure \ref{fig:xyfoc_ngc} illustrates the behaviour for the NGC (Figure \ref{fig:xyfoc_sgc} is similar, for the SGC).
The top panels show the data before the weighting by $/1f_{\text{noz,XYFOCAL}}$.
Some regions have either systematically lower-than-average (XFOCAL$\sim$-300, YFOCAL$\sim$-100; or extreme XFOCAL values) or higher-than-average (XFOCAL$\sim$-50, YFOCAL$\sim$50) SSR.
Our fitted model correctly reproduces that behaviour, as one can see from the red line in the side top panels, or in the bottom panels, which display the SSR after weighting by $1/f_{\text{noz,XYFOCAL}}$.

The total redshift failure weight $w_{\rm noz}$ applied on the data is the inverse product of $f_{\text{noz,pSN}}$ and $f_{\text{noz,XYFOCAL}}$.
To avoid double counting redshift failures, we weight each object by the median SN correction to perform the (XFOCAL,YFOCAL) fit (Eq.~(\ref{eq:zfailxyfoc})).

\subsection{Fibre collision and tiling completeness \label{sec:cp}}
When two or more targets are closer than the fibre collision radius (62 arcsec on the sky), they cannot not be spectroscopically observed within a single plate.
Those targets are said to `collide', and form what we call a `collision group' \citep[see ][for more details]{blanton03b,reid16}.
This effect has to be corrected in the analysis, as it artificially changes the clustering of the sample.
We weight each ELG with a valid fibre by the collision pair weight $w_{\rm cp}$ given by the number of targets over the number of valid fibres within each collision group.
Collided or not valid fibres are declared resolved when they lie in the same collision group as an ELG valid fibre \citep[see also ][]{mohammad20}.

The tiling completeness COMP\_BOSS is defined as the ratio of the number of resolved fibres to the number of targets in each sector, a sector being a region defined by a unique set of overlapping plates.
The tiling completeness is included in the randoms systematic weight $w_{\rm sys}$ and can be seen in Figure \ref{fig:tiling}.

\subsection{Systematics due to photometry}

Once corrected for systematics related to spectroscopic observations ($w_{\rm noz}$ and $w_{\rm cp}$), our $0.6<\zspec<1.1$ data sample still has (angular) imprints of the photometry used for target selection, that need to be corrected for.
Firstly, in regions with shallow imaging, higher photometric noise implies that more $\zspec<0.6$ objects than $\zspec>0.6$ objects enter our selection box in the $grz$-diagram, because of the density gradient in that $grz$-diagram; we thus expect to have less $0.6<\zspec<1.1$ objects in shallow imaging regions.
Other regions where we expect to have less $0.6<\zspec<1.1$ objects overall are regions with high Galactic extinction (because objects are dimmer) or regions with high stellar density (because each star is likely to blend with an ELG, which was not selected).

We include the following systematic photometric quantities as a source of systematics: the DECaLS imaging depth (\texttt{galdepth}, 5$\sigma$ detection limit for a galaxy with an exponential profile with a radius of 0.45 arcsec) and seeing (\texttt{psfsize}) for the three $grz$-bands, the stellar density (estimated from \textit{Gaia}/DR2), and the Galactic extinction, using E(B-V), dust temperature \citep{schlegel98}, and the HI column density \citep{lenz17,hi4pi16}.

\begin{figure}[h]
\begin{center}
\includegraphics[width=0.95\columnwidth]{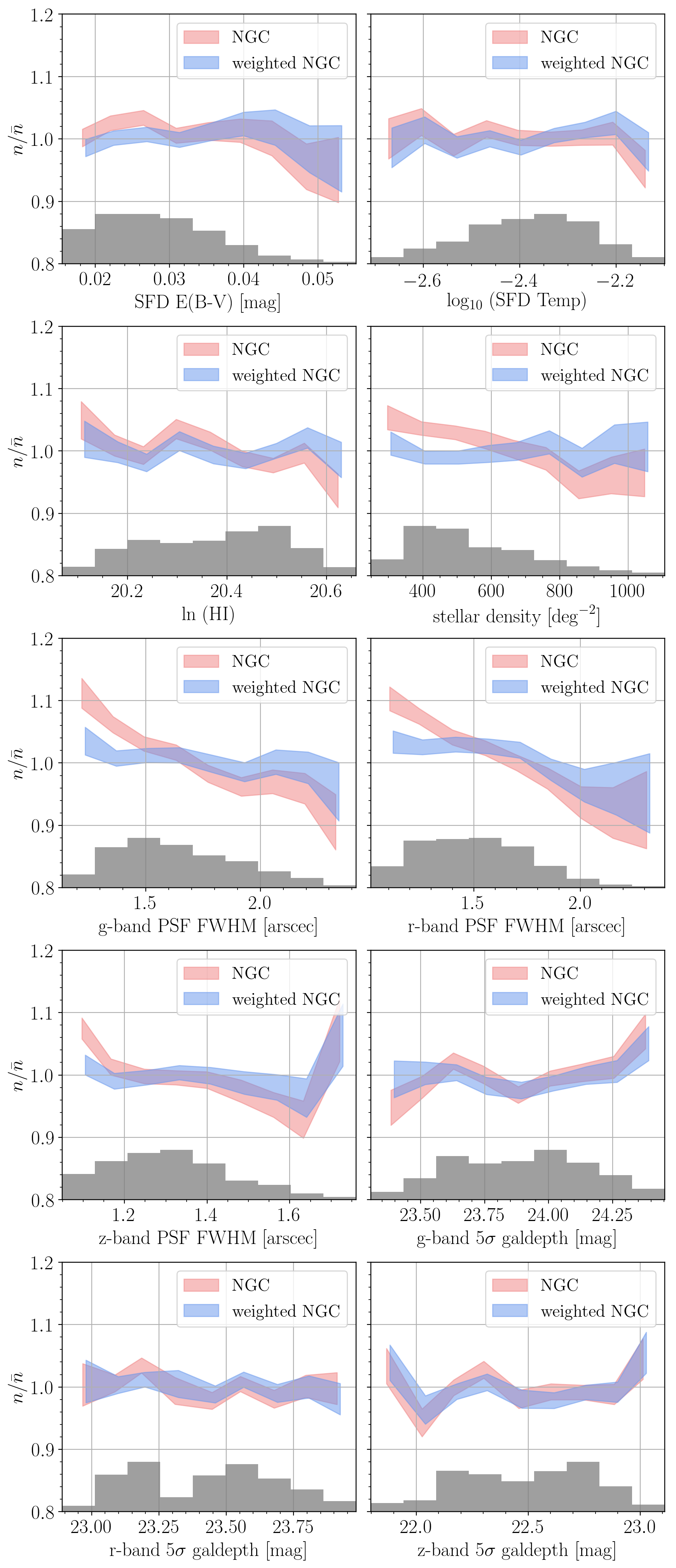}
\caption{Density fluctuations in the NGC for the $0.6<\zspec<1.1$ ELGs with a reliable $\zspec$, weighted by $w_{\rm noz} \cdot w_{\rm cp}$, before (red) and after (blue) applying the $w_{\rm sys}$ weights.
The systematics are: E(B-V) and dust temperature, HI column density, stellar density (from \textit{Gaia}/DR2), $grz$-band imaging seeing, $grz$-band imaging depth.
In each panel, we also display with the filled gray histogram the distribution of systematics values over the considered cap.}
\label{fig:sys_ngc}
\end{center}
\end{figure}

\begin{figure}[h]
\begin{center}
\includegraphics[width=0.95\columnwidth]{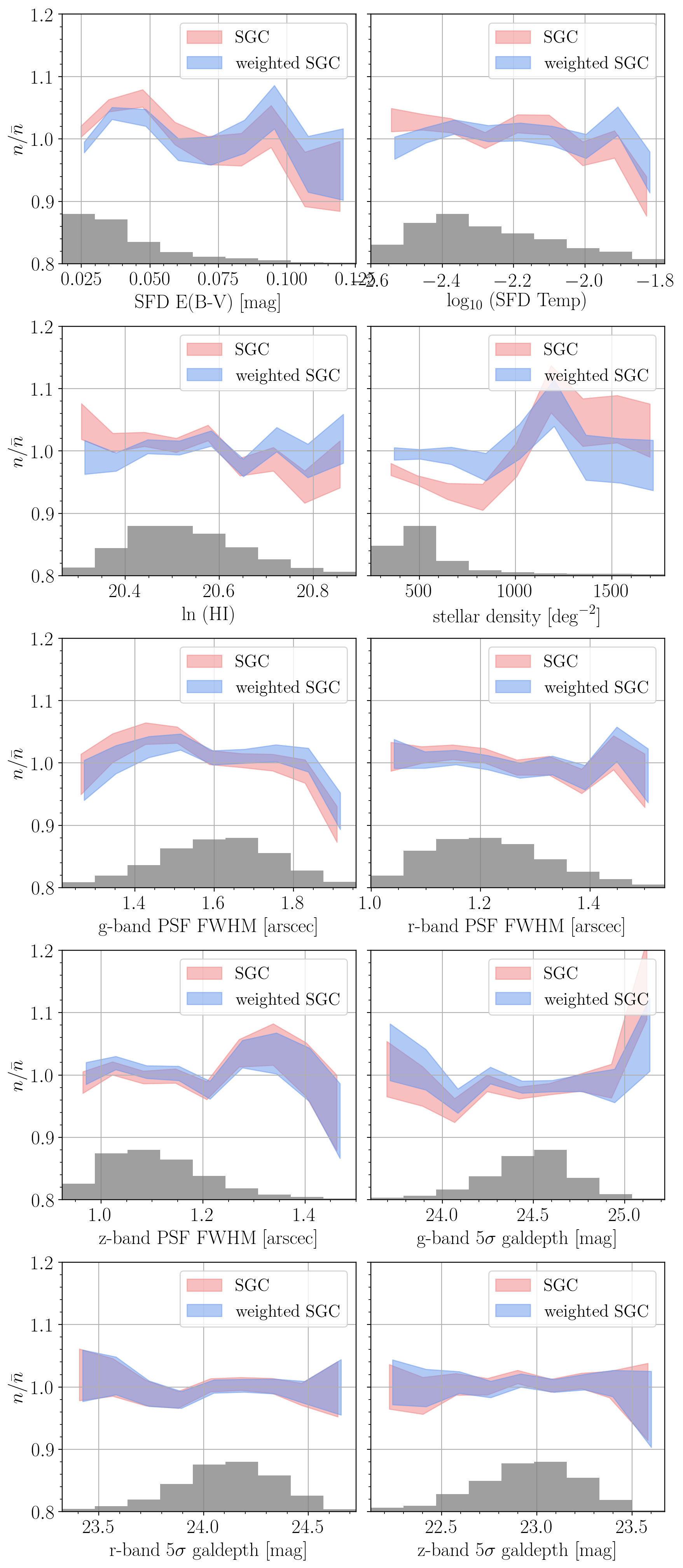}
\caption{Same as Figure \ref{fig:sys_ngc}, but for the SGC.
}
\label{fig:sys_sgc}
\end{center}
\end{figure}

To compute the $w_{\rm sys}$ weights to correct for systematics due to photometry, we first apply the veto masks both to our data and random samples.
We split the sky in \texttt{Healpix} pixels with \texttt{nside}=256 (area $\sim$ 0.05 deg$^2$).
For each pixel $p$, we firstly compute the median value $s_p$ for each photometric quantity.
Then, we compute $n_{\rm dat,p}$, the number of data weighted by $w_{\rm noz} \cdot w_{\rm cp}$, i.e. the number of $0.6<\zspec<1.1$ ELGs corrected for spectroscopic biases.
The number of randoms weighted by COMP\_BOSS, $n_{\rm ran,p}$, is obtained to derive the effective fractional area of each pixel.
For each chunk, we proceed to a multilinear fitting with minimising the $\chi^2_{\rm chunk}$ defined as:

\begin{equation}
\chi^2_{\rm chunk} = \sum_{p \in \; \rm P}
\left[
\frac{n_{\rm{dat},p} - n_{\rm{ran},p} \cdot ( \epsilon + \sum_{s \in S} c_s \cdot s_p)}{\sigma_{p}}
\right]^2,
\label{eq:sys}
\end{equation}
where $P$ is the list of the \texttt{Healpix} pixels inside the considered chunk, $S$ is the list of the photometric templates, $\sigma_p=\sqrt{n_{\rm{ran},p}}$ is the Poissonian error, and ($\epsilon$,$c_s$) are the fitted parameters.
We can then use the ($\epsilon$, $c_s$) fitted parameters to define the weight for each \texttt{Healpix} pixel $p$:

\begin{equation}
w_{\rm sys,p} = \frac{1}{\epsilon + \sum_{s \in S} c_s \cdot s_p}
\label{eq:wsys}
\end{equation}
Figures \ref{fig:sys_ngc} and \ref{fig:sys_sgc} display the dependency of the ELG density for each systematics $s$ before (red) and after (blue) applying the computed $w_{\rm sys}$, for the NGC and SGC, respectively.
We see that our computation reduces the density variations where they are the strongest, e.g. for \texttt{psfsize} or the stellar density in the NGC.

We refer the interested reader to \citet{kong20}, who find consistent results with a fully independent method.
Their approach, developed in the DESI context and tested on the eBOSS/ELG sample, consists in injecting fake, realistic sources in the imaging itself, running the \texttt{legacypipe} photometric pipeline on it, and then applying the target selection.
The strength of that approach is that it naturally accounts for any possible imaging systematics due to imaging.

\subsection{Weight normalisation}
The mean of photometric weights $w_{\rm sys}$ of all ELG targets is normalised to 1 in each chunk.
$w_{\rm noz}$ is then scaled such that the mean of the data completeness weights $w_{\rm sys} \cdot w_{\rm cp} \cdot w_{\rm noz}$ of ELGs with a reliable redshift or stars (the latter being assigned $w_{\rm noz}=1$) is equal to the mean of $w_{\rm sys}$ over all resolved fibers. Then targets with collided or invalid fibres are assigned $w_{\rm cp} = 0$.
Objects that have an unreliable redshift or stars are assigned $w_{\rm noz} = 0$.

\subsection{Random redshifts and weights \label{sec:randz}}

Once cut over the chunk footprint and the angular veto masks, the randoms have the same angular distribution as data.
We then need to attribute to the randoms redshifts with a similar radial distribution as the data.
We assign redshifts to randoms following the \textit{shuffled} scheme, i.e. picking up $\zspec$ values from the data, with a probability proportional to $w_{\rm noz} \cdot w_{\rm cp} \cdot w_{\rm sys}$, so that the weighted distributions of data and randoms match.

\begin{figure}
\begin{center}
\includegraphics[width=0.95\columnwidth]{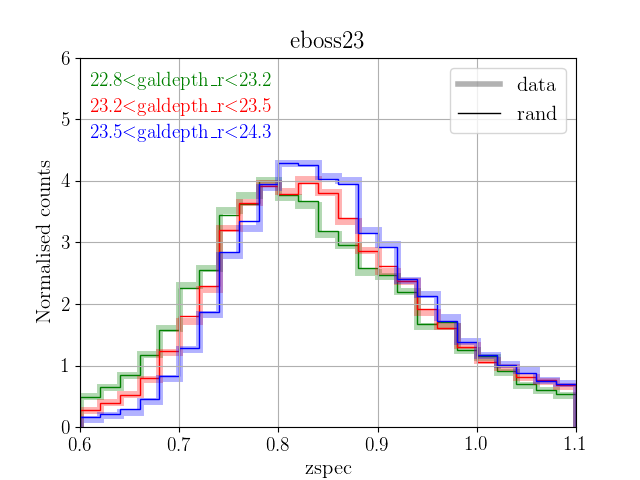}
\caption{Illustration of the dependency of redshift distribution on imaging depth for the \texttt{eboss23} chunk, where the dependency is strong.
Our randoms (thin lines) faithfully reproduce the trend of the data (thick lines).}
\label{fig:nzdepth}
\end{center}
\end{figure}

However, we need to account for another effect.
The ELG data $n(z)$ depends on the depth of the imaging used for target selection (markedly for \texttt{eboss23}, but also in the SGC), with $n(z)$ having more $\zspec<0.8$ ELGs in shallow imaging regions.
Figure \ref{fig:nzdepth} illustrates that effect for the $r$-band imaging in \texttt{eboss23}, where the sample is split in three bins of $r$-band imaging depth.
This implies an angular-radial relation that needs to be accounted for in the randoms.

To account for this effect of depth on the target selection process, we split each chunk in  three subregions of approximately constant imaging depth and apply the \textit{shuffled} scheme in each subregion.
We define the three subregions with modelling the $n(z)$ as a simple function of flux limits.
We first define, at any position in the chunk, $f_{grz}$, a combined $grz$-band imaging depth that correlates at best with the data $\zspec$.
We define $f_{grz} = \epsilon + c_g f_g + c_r f_r + c_z f_z$, a linear combination of $f_g, f_r, f_z$, the 5$\sigma$ flux detection limits of the imaging at the position of an ELG in the $g$-, $r$-, $z$-bands.
The ($\epsilon$, $c_g$, $c_r$, $c_z$) coefficients are the fitted with minimising:
\begin{equation}
\chi^2_{grz} = \sum_{i=1}^{N_g}
\left[
z_{\text{spec},i} -(\epsilon + c_g f_g^i + c_r f_r^i + c_z f_z^i)
\right]^2 \times w_{\rm noz}^i \cdot w_{\rm cp}^i \cdot w_{\rm sys}^i,
\label{eq:nzdepth}
\end{equation}
where the sum is over the $N_g$ ELGs of the chunk.
We then bin the randoms in three bins of $f_{grz}$, hence defining the three subregions of approximately constant depth imaging; the data are binned with the same three subregions.
For a random with a $f_{grz}$ value, we pick a redshift from the data $\zspec$ from the corresponding $f_{grz}$ bin, with a probability proportional to $w_{\rm noz} \cdot w_{\rm cp} \cdot w_{\rm sys}$.
That approach allows us to reproduce this dependency in the randoms redshifts, as can be seen in Figure \ref{fig:nzdepth}, where the randoms weighted $n(z)$ closely follows that of the data when splitting by $r$-band imaging depth.

For randoms, weights are defined as follows: $w_{\rm sys}$ is the tiling completeness COMP\_BOSS, and $w_{\rm noz} = w_{\rm cp} = 1$.
Then, random weights are normalised to ensure that the sum of weighted data over the sum of weighted randoms is
the same in each \texttt{chunk\_z}.

Using the \textit{shuffled} scheme introduces a radial integral constraint \citep[RIC,][]{de-mattia19}, which is particularly important for this sample, as the random $n(z)$ is tuned to the data $n(z)$ in small chunks.
We correct for that effect with using the formalism introduced in \citet{de-mattia19}.
\citet{zhao20} and \citet{tamone20} study the impact of that correction for the different multipoles, for the mocks and the data, respectively.
The monopole is marginally changed, whereas the quadrupole and the hexadecapole are significantly changed.

Lastly, we remove 163 randoms belonging to tiny sectors where there are no data with a reliable $\zspec$, which is equivalent to restricting to sectors with \texttt{COMP\_BOSS}$\geq$0.5 and SSR$\geq$0.

\subsection{FKP and redshift distribution}

\begin{figure}
\begin{center}
\includegraphics[width=0.95\columnwidth]{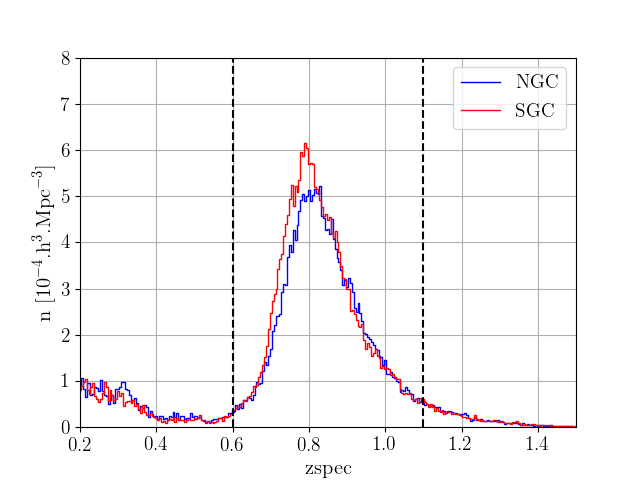}
\caption{Number density of ELGs in the eBOSS survey. The vertical dashed lines indicate the redshift range used in our clustering measurement.}
\label{fig:nz}
\end{center}
\end{figure}

The redshift distribution of our ELG sample, split by NGC and SGC, is displayed in Figure \ref{fig:nz}.
The effective redshift of our sample is $z_{\rm eff} = 0.845$.
We use the fiducial eBOSS DR16 cosmology (reported in Table \ref{tab:cosmo}) to derive the comoving number density.

\begin{table}
\centering
\caption{Different cosmologies and redshift used in this paper. $h$ is defined such that $H_0 = 100 \times h$ \kms Mpc$^{-1}$. All cosmologies are flat $\Lambda$CDM, hence $\Omega_\Lambda = 1 - \Omega_{\rm m}$. The BAO fits in Section \ref{sec:bao} are performed with the `DR16 Fiducial' cosmology.}
\begin{tabular}{lccc}
\hline
\hline
 & DR16 Fiducial & OuterRim & EZmocks\\
\hline
$h$                     & 0.676 & 0.71      & 0.6777    \\
$\Omega_{\rm m}$        & 0.31  & 0.26479   & 0.307115  \\
$\Omega_{\rm b}h^2$     & 0.022 & 0.02258   & 0.02214   \\
$\sigma_8$              & 0.8   & 0.8       & 0.8225    \\
$n_{\rm s}$             & 0.97  & 0.963     & 0.9611    \\
$\Sigma m_\nu$ [eV]     & 0.06  & 0         & 0         \\
\hline
redshift               & $z_{\rm eff} = 0.845$ & $z_{\rm snap} = 0.865$ & $z_{\rm eff} = 0.845$\\
\hline
\end{tabular}
\label{tab:cosmo}
\end{table}

As in previous BOSS/eBOSS analyses \citep[e.g.][]{anderson14,alam17,ata18}, we define inverse-variance $w_{\rm FKP}$ weights to be applied to data and randoms.
We define $w_{\rm FKP} = 1/(1+n(z) \cdot P_0)$ \citep{feldman94}, where $P_0 = 4000 h^{-3}$ Mpc$^3$ is the amplitude of the power spectrum at the $k$ scale at which the FKP-weights minimise the variance of the measurement \citep{font-ribera14a}.
Since n(z) varies with the local clustering, the $w_{\rm FKP}$ weights tend to upweight (resp. down-weight) underdensities (resp. overdensities). We did verify that the induced systematic bias is small enough for our analysis.

\subsection{Effects of weights on the monopole}
\begin{figure}
\begin{center}
\begin{tabular}{c}
\includegraphics[width=0.95\columnwidth]{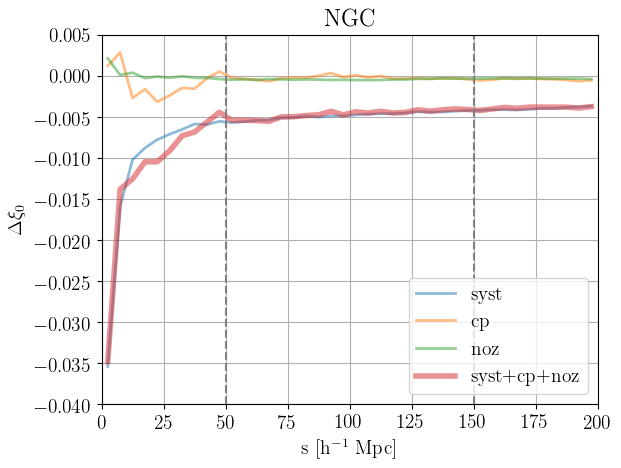}\\ 
\\[20pt]
\includegraphics[width=0.95\columnwidth]{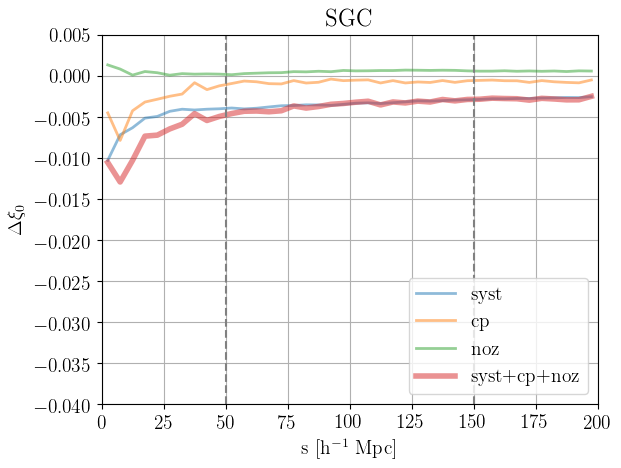}\\
\end{tabular}
\caption{Effect of the weights on the clustering for the NGC (top panel) and the SGC (bottom panel). The vertical lines show the BAO fitting range in this paper.}
\label{fig:xi0weights}
\end{center}
\end{figure}

We display in Figure \ref{fig:xi0weights} how the weights computed in the previous sections change the clustering of the ELG sample.
As expected \citep[see e.g. ][]{ross17,ata18}, the $w_{\text{sys}}$ weights have by far the strongest impact on the clustering.
We notice that the $w_{\text{cp}}$ weights have an impact at all scales in the SGC and decreasing the clustering: a possible interpretation is the ELG SGC chunk geometry, noticeably \texttt{eboss21} with its small area.
Close pairs should have been missed preferentially around the edges and there are more edges because of the small footprint.
Lastly, the $w_{\text{noz}}$ weights have a marginal impact on the clustering.

\section{Mock catalogues} \label{sec:mocks}

In order to validate and perform our BAO fitting method, we rely on two sets of mock catalogues.
The cosmology of each set of mock is reported in Table \ref{tab:cosmo}.
We refer the reader to \citet{de-mattia20} for more details on those both sets of mocks.

\subsection{Accurate N-body Sky-cut OuterRim mocks}

The first set of mock catalogues used in the subsequent BAO analysis are the 6 Sky-cut OuterRim mocks, described in \citet{de-mattia20}.
The starting product is the OuterRim simulation \citep{heitmann19}, which is one of the largest high-resolution N-body simulations to date, as it contains 10,240$^3$ particles with a mass of $1.85 \cdot 10^9$ $h^{-1}$ M$_\odot$ over a volume of (3000 $h^{-1}$ Mpc)$^3$.
\citet{avila20} have extracted from the OuterRim simulation the snapshot at $z_{\rm snap} = 0.865$ and have produced accurate mocks, which faithfully reproduce the DR16 ELG data sample small-scale clustering, using the Halo Occupation Distribution modelling motivated by \citet{gonzalez-perez18}.
From those \citet{avila20} mocks, the Sky-cut OuterRim mocks are generated, by cutting the eBOSS/ELG footprint, applying the veto masks, and reproducing the data $n(z)$ distribution and accounting for the $n(z)$ dependence with the imaging depth.

\subsection{Approximate EZmocks} \label{sec:ezmocks}

The second set of mocks consists of the 1000 EZmocks realisations presented in \citet{zhao20}.
The EZmocks are using the Zel'dovich approximation \citep{zeldovich70} to generate a density field and populate galaxies according to the desired tracer bias.
As for the Sky-cut OuterRim mocks, those EZmocks are cut according to the eBOSS/ELG footprint, have the veto masks applied, reproduce the data $n(z)$ distribution, and account for the dependence with the imaging depth.

Additionally, we build another set of 1000 EZmocks, where we include the observational systematics present in the data \citep[see also][]{de-mattia20}: we implement the spectroscopic systematics (fibre collision and redshift failures) and the angular systematics (mocks are produced at a density higher than the ELG one, and are then trimmed according to a smoothed map of the data observed density, thus accounting for possibly unknown angular photometric systematics). 
For each mock, we then compute the weighting scheme as we do for the data.
We remark that, since weights are recomputed on each mock, the noise in the weight calculation due to shot noise and cosmic variance is automatically propagated to the final cosmological parameters. 

Those EZmocks with observational systematics are the ones used in Section \ref{sec:bao}, in particular to estimate the covariance matrices.
The set of EZmocks without systematics are only used in Section \ref{sec:mockres}, when comparing to the OuterRim mocks which have no systematics included.

\section{The model and fitting methodology} \label{sec:bao}

\subsection{The model}

We measure spherically averaged BAO measurements using the 2-point correlation function.
Our methodology closely follows that described in \citet{anderson14,ross17,ata18} and references therein, to which we refer for more details.

We first compute $\xi(s,\mu)$, the redshift-space 2D correlation function as a function of $s$, the separation vector in redshift-space and $\mu$ the cosine of the angle between $s$ and the line-of-sight direction.
We use the \citet{landy93} estimator:

\begin{equation}
\xi(s,\mu) = \frac{DD(s,\mu) - 2DR(s,\mu) + RR(s,\mu)}{RR(s,\mu)},
\label{eq:ls}
\end{equation}
where $DD$, $DR$, and $RR$ are the normalised number of data-data, data-random, random-random pairs with a separation of $s$ and an orientation of $\mu$\footnote{The pair-counting is done using the `DR16 Fiducial`, `OuterRim', and `EZmocks' cosmology for the data, the OuterRim mocks, and the EZmocks, respectively.}.
We then compute the monopole correlation function $\xi_0(s)$, i.e. the first Legendre multipole with:
\begin{equation}
\xi_l(s) =  \frac{2l+1}{2} \int_{-1}^{1} L_l(\mu) \xi(s,\mu) d\mu \;\; \text{for} \;\; l=0,
\label{eq:xi0}
\end{equation}
where $L_l(\mu)$ is the $l^{th}$-order (0$^{th}$ here) Legendre polynomial.
 
We measure the difference in the BAO location between our clustering measurement and that expected in our fiducial cosmology, which can mostly come either from a difference in projection or from the difference between the BAO position in the true intrinsic primordial power spectrum and that in the model, with the multiplicative shift depending on the ratio $r_{\rm drag}/r^{\rm fid}_{\rm drag}$, where $r_{\rm drag}$ is the comoving sound horizon at $z=z_{\rm drag}$, the redshift at which the baryon-drag optical depth equals unity \citep{hu96}.
If we define the spherically averaged distance $D_V(z) = \left[D_M^2(z) \cdot czH(z)^{-1}\right]^{1/3}$ as a combination of the Hubble parameter $H(z)$ and the comoving angular diameter distance $D_M(z)$, we can express the offset between the observed BAO location and our template as:
\begin{equation}
\alpha =  \frac{D_V(z)r^{\rm fid}_{\rm drag}}{D^{\rm fid}_V(z)r_{\rm drag}}.
\label{eq:alpha}
\end{equation}
 
 Once we have our measurement of $\alpha$, it can be converted to an angular location of the BAO, a dimensionless quantity that is independent of cosmology:
 \begin{equation}
     \frac{D_V(z_{\rm eff}=0.845)}{r_{\rm drag}} = \alpha \frac{D^{\rm fid}_V(z_{\rm eff}=0.845)}{r^{\rm fid}_{\rm drag}}.
     \label{eq:dvrd}
 \end{equation}
For our fiducial cosmology (`DR16 Fiducial' in Table \ref{tab:cosmo}), $r^{\rm fid}_{\rm drag}=147.77$ Mpc and $D^{\rm fid}_V(z_{\rm eff}=0.845) = 2746.8$ Mpc.

 We generate a template BAO feature using the linear power spectrum, $P_{\rm lin}(k)$, obtained from {\sc Camb}\footnote{\niceurl{https://camb.info/}} \citep{lewis00,howlett12} and a `no-wiggle' $P_{\rm nw}(k)$ obtained from the \citet{eisenstein98} fitting formulae\footnote{In order to best-match the broadband shape of the linear power spectrum, we use $n_s = 0.963$, to be compared to 0.97 when generating the full linear power spectrum from {\sc camb}.
 This linear power spectrum is same as used for BOSS and eBOSS galaxy analyses since DR11.}, both using our fiducial cosmology (except where otherwise noted). 

Given $P_{\rm lin}(k)$ and $P_{\rm nw}(k)$, we account for redshift-space distortion (RSD) and non-linear BAO damping via
\begin{equation}
P(k,\mu) = C^2(k,\mu,\Sigma_s)\left((P_{\rm lin}-P_{\rm nw})e^{-k^2\sigma_v^2}+P_{\rm nw}\right),
\end{equation}
where
\begin{equation}
\sigma^2_v = (1-\mu^2)\Sigma^2_{\perp}/2+\mu^2\Sigma^2_{||}/2,
\end{equation}

\begin{equation}
C(k,\mu,\Sigma_s) = \frac{1+\mu^2\beta(1-S(k))}{(1+k^2\mu^2\Sigma^2_s/2)}.
\label{eq:Csk}
\end{equation}
$S(k)$ is the smoothing applied in reconstruction: $S(k) = e^{-k^2\Sigma_r^2/2}$ and $\Sigma_r = 15 h^{-1}$Mpc for the reconstruction applied to the eBOSS ELG sample (see Section \ref{sec:recon}); $S(k)=0$ for pre-reconstruction. This matches the implementation of \cite{ross17}, which was motivated by \cite{Seo16}.
For our fiducial analysis, we fix $\beta=0.593$ and $\Sigma_s = 3h^{-1}$Mpc.
Given this is a spherically averaged analysis that does not consider how the signal changes with respect to the line of sight, we expect these parameters to have no significant effect. 
We use $\Sigma_{\perp} = 3 h^{-1}$Mpc and $\Sigma_{||} = 5 h^{-1}$Mpc for post-reconstruction results and $\Sigma_{||}= 10 h^{-1}$Mpc and $\Sigma_{\perp}= 6 h^{-1}$Mpc for pre-reconstruction.
We discuss these choices for the damping parameters in further detail when discussing results achieved from mock catalogues in Section \ref{sec:mockres}.

In order to produce our spherically averaged BAO template in the configuration space, $\xi_{\rm temp}$, we use the Fourier transform of $P_0(k) = \int {\rm d}\mu P(k,\mu)$.
We then fit the model:
\begin{equation}
    \xi_{\rm mod}(s,\alpha) = B\xi_{\rm temp}(s\alpha) + A_0 + A_1/s+A_2/s^2.
\end{equation}
For $B$, we use a Gaussian prior of width 0.4 around $B/B_{\rm fit}$, where $B_{\rm fit}$ is the value of $B$ one obtains from the first measurement bin in the $\xi_0$ data vector ($50 < s < 55 h^{-1}$Mpc in our fiducial case) when fixing $A_N=0$. 

In addition to damping the BAO oscillations, non-linear evolution effects are also expected to cause small shifts (of order 0.5 percent) in the BAO position \citep{padmanabhan09}, which should have a small cosmological dependence (e.g., the size of the shift is likely dependent on $\sigma_8$).
Reconstruction has been demonstrated to reverse such effects and we will discuss any residual systematic uncertainty in Section \ref{sec:mockres}.

\subsection{Parameter estimation}

As in \citet{ata18}, we assume the likelihood distribution, ${\cal L}$, of any parameter (or vector of parameters), $p$, of interest is a multi-variate Gaussian:  
\begin{equation}
{\cal L}(p) \propto e^{-\chi^2(p)/2}.
\end{equation}
The $\chi^2$ is given by the standard definition
\begin{equation}
\chi^2 = {\bf D}{\sf C}^{-1}{\bf D}^{T},
\end{equation}
where ${\sf C}$ represents the covariance matrix of the measured correlation function and ${\bf D}$ is the difference between the data and model vectors, when model parameter $p$ is used.
Our DR16 fiducial cosmology (Table \ref{tab:cosmo}) is always used in the fits.
We assume flat priors on all model parameters, unless otherwise noted.
Our fitting range is $50 < s < 150 h^{-1}$Mpc, with using $5 h^{-1}$Mpc bins for our fiducial $\xi(s)$ results.
These choices match those applied in \cite{ross17}, which were found to be appropriate for post-reconstruction data.

Similar to previous analyses (e.g., \citealt{ata18}), we obtain $\chi^2(\alpha)$ by finding the value of the nuisance parameters that minimises $\chi^2(\alpha)$.
We do this on a grid of spacing 0.001 in the range $0.8 < \alpha < 1.2$.
We define a `detection' as there being a $\Delta \chi^2=1$ region on both sides of the minimum $\chi^2$. To report the results we use the Gaussian approximation that the uncertainty on the measurement as half of the width of this $\Delta \chi^2=1$ region and the maximum likelihood its mean.
We recommend use of the full $\chi^2(\alpha)$ result for testing cosmological models, rather than this Gaussian approximation. This will be made publicly available after this work is accepted for publication.

In order to estimate covariance matrices, we use the 1000 approximate EZmocks with systematics included, which mimick our ELG sample (see Section \ref{sec:ezmocks}). 
The noise from the finite number of mock realisations requires some corrections to the $\chi^2$ values, the width of the likelihood distribution, and the standard deviation of any parameter determined from the same set of mocks used to define the covariance matrix.
These factors are defined in \citet{hartlap07}, \citet{dodelson13} and \citet{percival14}; we apply the factors in the same way as in, e.g., \citet{anderson14,ata18}.
For our fiducial $\xi(s)$ results, we use 1000 mocks and 20 measurement bins for each NGC and SGC regions.
Thus, the number of mock realisations is much larger than the number of measurement bins, implying the finite number of mocks has less than a 2 percent effect on our uncertainty estimates. 

\subsection{Reconstruction}
\label{sec:recon}

BAO measurements can be improved by applying `reconstruction' techniques that partially remove non-linear effects on the BAO feature observed in 2-point clustering measurements \citep{eisenstein07}.
We apply the reconstruction method presented in \citet{burden15} and further described in \citet{bautista18}.
We use the case where RSD are removed and three iterations are applied.
We assume the ELG sample has a bias of 1.4 (approximately correct for our sample and fiducial cosmology), and we assume the growth rate $f$=0.82.
As in previous studies, we use a smoothing scale of 15$h^{-1}$Mpc. The particular parameters applied are not expected to bias the results (see, e.g., \citealt{vargas-magana18}).

\subsection{Comparing clustering in data and mocks}

\begin{figure}
\centering
\includegraphics[width=0.45\textwidth]{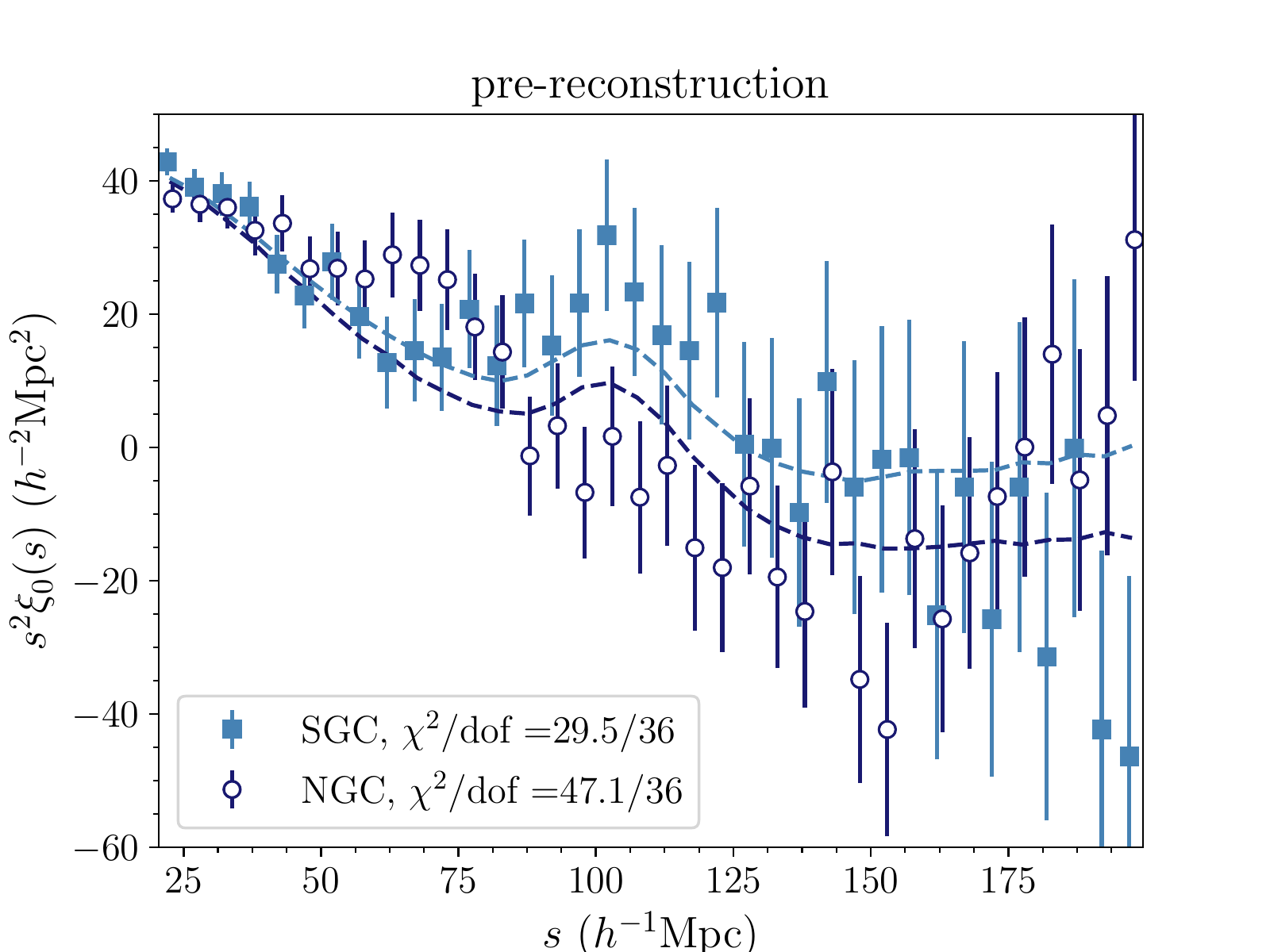}
\includegraphics[width=0.45\textwidth]{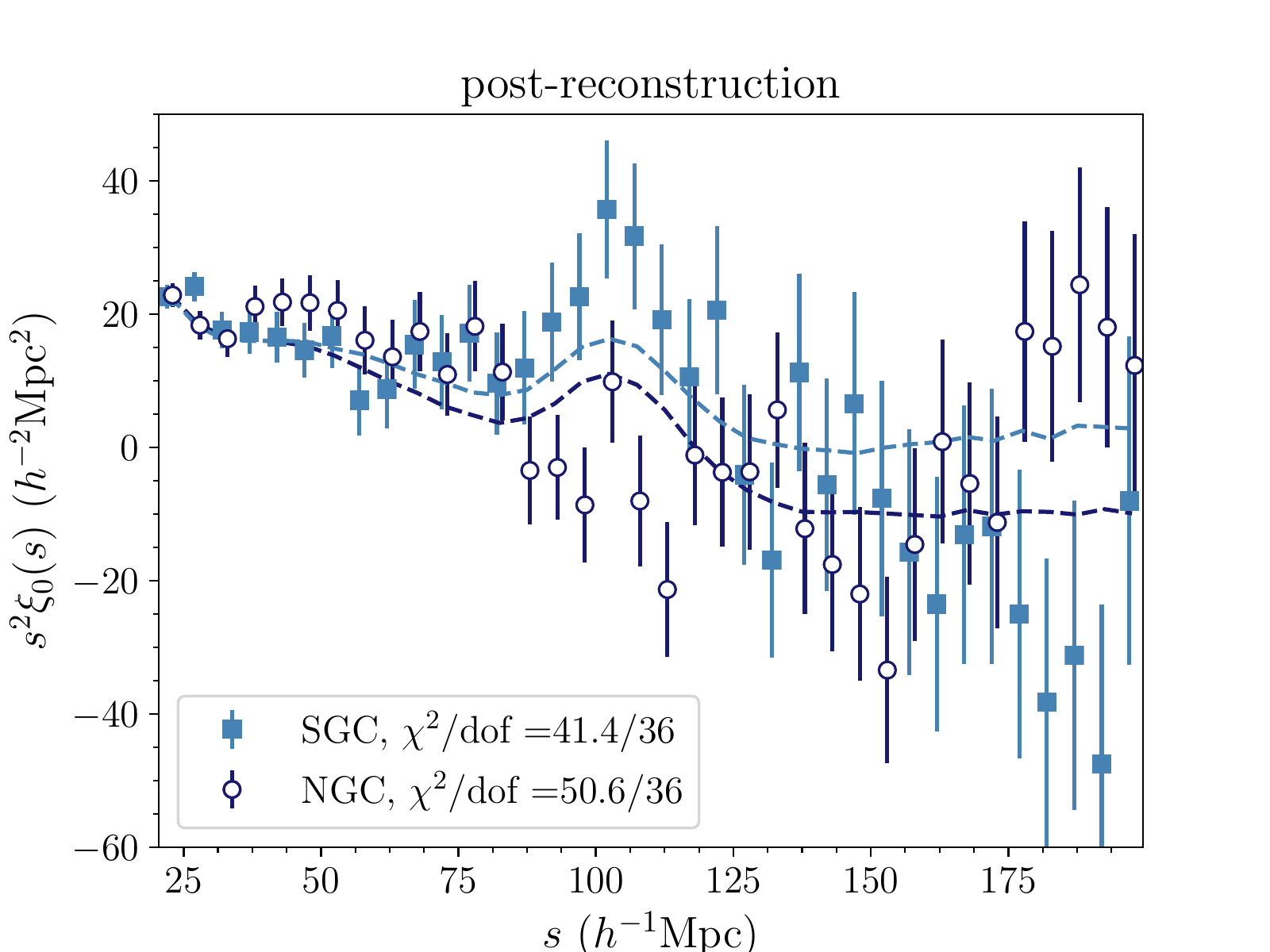}
\includegraphics[width=0.45\textwidth]{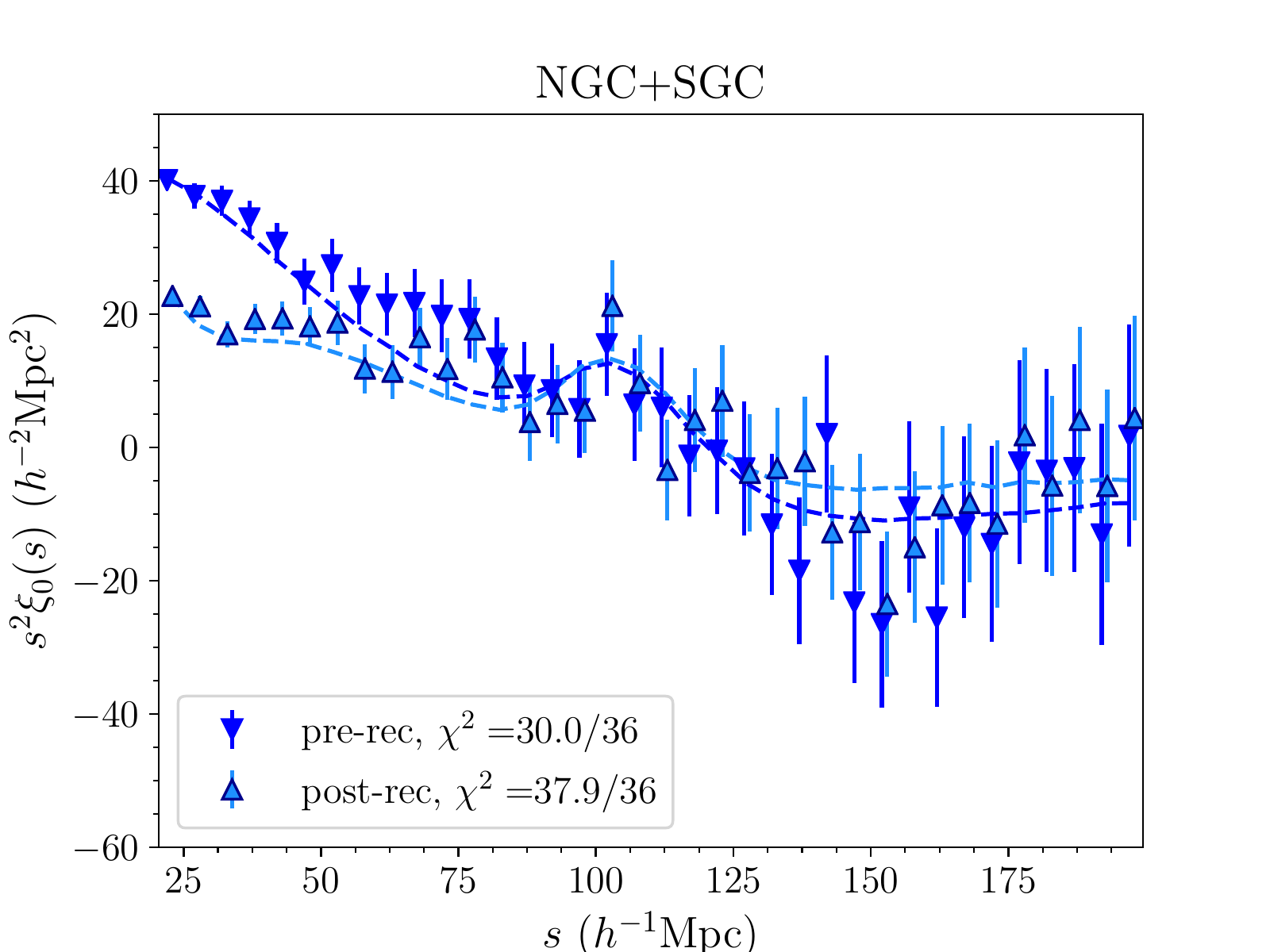}
\caption{The measured DR16 correlation function of data ELGs (points with error bars) compared to the mean of the EZmocks (dashed lines). NGC and SGC are compared pre- and post-reconstruction in the top two panels. The bottom panel compares the NGC+SGC combination for both.}
\label{fig:xicompez}
\end{figure}

In Figure \ref{fig:xicompez}, we display the spherically-averaged redshift-space correlation functions we use for BAO measurements, compared to the mean of the EZmocks.
The $\chi^2$/dof between the data and the mocks for the comparison are labelled in each panel of the figures.
While we do expect these to be of order 1, some deviation is expected given that the EZmocks are approximate and the fiducial EZmock cosmology is expected to be somewhat different than the true cosmology (in unknown directions, of course).

The pre-reconstruction results are shown in the top panel of Figure \ref{fig:xicompez}.
Immediately noticeable is the fact that the large-scale clustering amplitude is expected to be lower in the NGC compared to the SGC, and the results for the data are consistent with this expectation. The underlying HOD applied to the EZmocks is the same in both hemispheres.
The difference in large-scale clustering amplitude is due to the fact that the $n(z)$ in the NGC is strongly dependent on the imaging depth and our treatment of this imparts an extra radial integral constraint.
In the NGC, we also notice an excess of clustering at around 60 $h^{-1}$Mpc; our only potential explanation for this is that it is a statistical fluctuation, as the overall agreement between the mocks and data is reasonable ($\chi^2/{\rm dof} = 47.1/36$).
We notice an apparently strong BAO feature in the SGC data and no such feature in the NGC data.

The post-reconstruction results are shown in the middle panel of Figure \ref{fig:xicompez}.
The apparent BAO feature remains strong in the SGC and missing from the NGC data.
The pre-reconstruction excess at around 60 $h^{-1}$Mpc in the NGC result has mostly been removed post-reconstruction, though the overall agreement has gotten slightly worse ($\chi^2/{\rm dof} = 50.6/36$).

In the bottom-panel of Figure \ref{fig:xicompez}, we compare the inverse-variance (based on the diagonal of the covariance matrix) weighted combination of the NGC and SGC to the mean of the EZmocks weighted in the same way.
This demonstrates that the full sample agrees well with our expectations, over a range of scales $20 < s < 200 h^{-1}$Mpc that is significantly wider than we use for our BAO fits.
However, given the differences between the NGC and SGC shown in the top two panels, we will fit the NGC and SGC separately and combine their likelihoods in order to obtain our BAO results.

The fact that the EZmocks reproduce the clustering of the eBOSS DR16 ELG sample, including the differences between the NGC and SGC, suggest that they will provide a good covariance matrix for fitting the data. Further, the results suggest that applying our BAO fitting methodology to the EZmocks will provide a reasonably approximate statistical sample to interpret our fit to the data.

\subsection{Fitting mock catalogues} \label{sec:mockres}

In this section, we present tests of BAO fitting methodology on mocks.
We focus mostly on the post-reconstruction results.
We will first investigate the results obtained from the mean of the EZ and OuterRim ELG mocks and then consider the results obtained from individual EZmock realisations.

As detailed in Section 8.3 of \citet{beutler17}, approximate mocks may not provide as sharp a BAO feature as expected, (e.g., due to grid effects) and one may wish to use N-body mocks to probe the expected signal strength.
For this reason, BOSS DR12 used damping parameters motivated by the N-body results of \cite{Seo16}.
Here, we use the Sky-cut OuterRim ELG mocks as N-body mock representing our expectations for the ELG sample.

\begin{table}
\centering
\caption{Tests of BAO fits on the mean of ELG mocks. We quote the difference between the obtained $\alpha$ and that expected given the cosmology of the mock, $\alpha_{\rm exp}$. For the EZmocks, $\alpha_{\rm exp} = 1.000$ and for OuterRim $\alpha_{\rm exp} = 0.942$. All results use the EZmock covariance matrices and the quoted uncertainty is for one realisation (thus, the one should divide the uncertainty from the mean of the EZmocks by $\sqrt{1000}$ in order to compare to the total uncertainty). The $\chi^2$ values for a given set of mocks are included only to allow one to determine the relative goodness-of-fit.}
\begin{tabular}{lcc}
\hline
\hline
case  & $\alpha-\alpha_{\rm exp}$ & $\chi^2$\\
\hline
OuterRim mocks, post reconstruction:\\
$\Sigma_{\perp,||} = 3,5$ & $0.000\pm0.025$ & 0.36 \\
$\Sigma_{\perp,||} = 4,7$ & $0.000\pm0.026$ & 0.50 \\
EZmocks, post reconstruction:\\
$\Sigma_{\perp,||} = 3,5$ & $0.007\pm0.038$ & 0.23\\
$\Sigma_{\perp,||} = 4,7$ & $0.007\pm0.040$ & 0.11\\
$\Sigma_{\perp,||} = 5,8.5$ & $0.007\pm0.042$ & 0.10\\
EZmocks, post reconstruction, no sys:\\
$\Sigma_{\perp,||} = 3,5$ & $0.005\pm0.038$ & 0.08\\
$\Sigma_{\perp,||} = 4,7$ & $0.005\pm0.040$ & 0.04\\
$\Sigma_{\perp,||} = 5,8.5$ & $0.006\pm0.042$ & 0.09\\
EZmocks, pre reconstruction:\\
fiducial & $0.009\pm0.055$ & 0.11\\
\hline
\hline
\label{tab:baomockmean}
\end{tabular}
\end{table}

\begin{figure}
\begin{center}
\includegraphics[width=0.45\textwidth]{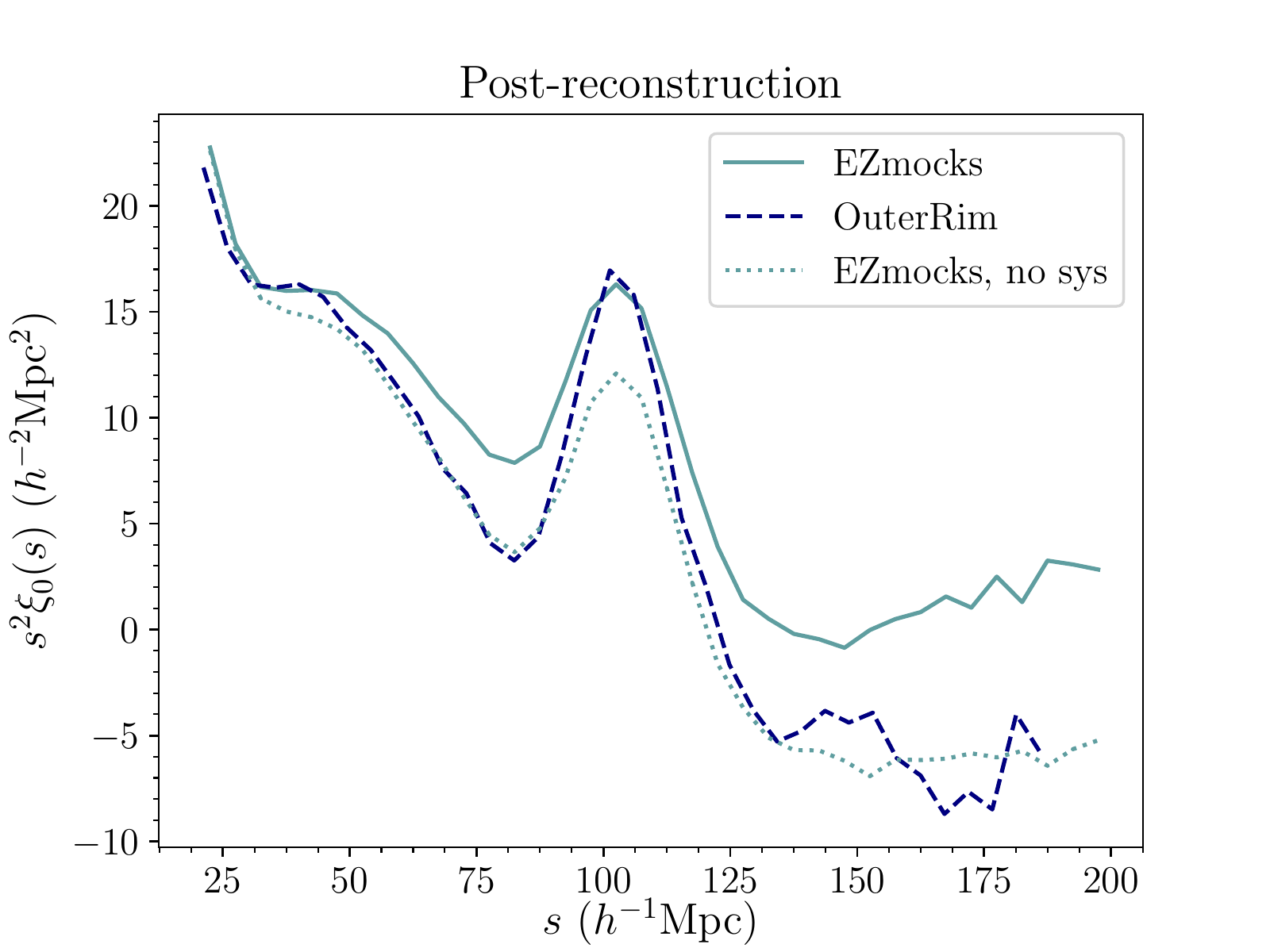} 
\caption{Comparison of the mean of the ELG EZ and OuterRim mocks in the SGC region. The OuterRim result has its $s$ values scaled by $\alpha=0.942$ in order to account for the difference in fiducial cosmologies.}
\label{fig:ezor}
\end{center}
\end{figure}

Our tests on the OuterRim mocks predict a significantly stronger BAO feature than the EZmocks.
Figure \ref{fig:ezor} displays the mean of the post-reconstruction EZ and OuterRim mocks in the SGC region.
The results for the EZmocks are shown with and without systematics imparted (the OuterRim mocks have no systematics imparted).
The broad-band shapes are in good agreement when there are no systematics, but the BAO feature is significantly sharper for the OuterRim mocks.
When systematic fluctuations are imparted, the broad-band amplitude is increased, but the sharpness of the BAO appears similar.

We investigate this further by fitting these mean $\xi_0$ with varying damping scales.
The results are presented in Table \ref{tab:baomockmean}.
For each case, we use the covariance matrix of the EZmocks with systematic fluctuations.
When systematic fluctuations are added to the EZmocks, the uncertainty that we obtain does not change (at the level of precision we quote); this indicates that indeed the BAO signal is nearly unaffected by the systematic fluctuations.
These uncertainties are 50 percent greater than those obtained from the OuterRim mocks.
Relatedly, we find the OuterRim mocks prefer smaller damping parameters than the EZmocks.
The OuterRim mocks are well-fit by damping parameters $\Sigma_{\perp},\Sigma_{||} = 3,5 h^{-1}$Mpc and we adopt these as our fiducial parameters to use for the data.
Importantly, it is the observed BAO signal that strongly impacts the fit precision, rather than the signal assumed by the model \citep[see for instance][]{hinton20}, i.e., the derived precision is only weakly dependent on the assumed $\Sigma_{\perp},\Sigma_{||}$. This is illustrated by the fact that the greatest variation in the uncertainty that is obtained when varying the damping parameters is only 10 percent (when changing from $3,5 h^{-1}$Mpc to $5,8.5 h^{-1}$Mpc), to be compared to the 50 percent variation found above.
The accuracy of the measurement is unaffected by this modelling choice, as $\alpha - \alpha_{\rm exp}$ changes by only 0.001.

The BAO measurement for the mean of the EZmocks is biased high, and given there are 1000 EZmocks, the significance is $>5\sigma$ for the mocks with systematic fluctuations.
However, compared to the precision we achieve on the data, it is less than 0.25$\sigma$ and thus not significant. 
Further, our results on the OuterRim simulation are unbiased, so it is unclear if it is our methodology or the nature of the approximate EZmocks causing the bias (especially given the same modelling techniques achieved unbiased results in the past). Some of the the shift can be attributed to the systematic fluctuations, as there is a 0.2 percent shift in $\alpha$ when the fluctuations are added.

Given that we expect the BAO signal to be stronger in the data than in the EZmocks, we therefore expect the uncertainty we achieve on the data to be better than the typical EZmock and closer to the OuterRim result.
Even so, studying the distribution of mock results is an important validation of the methodology and allows comparisons to other ELG analyses.
Given the strength of the BAO feature in the mean of the EZmocks, we use $\Sigma_{\perp,||} =  4,7h^{-1}$Mpc as the fiducial choice for fitting individual EZmock realisations.

\begin{figure}
\begin{center}
\includegraphics[width=0.45\textwidth]{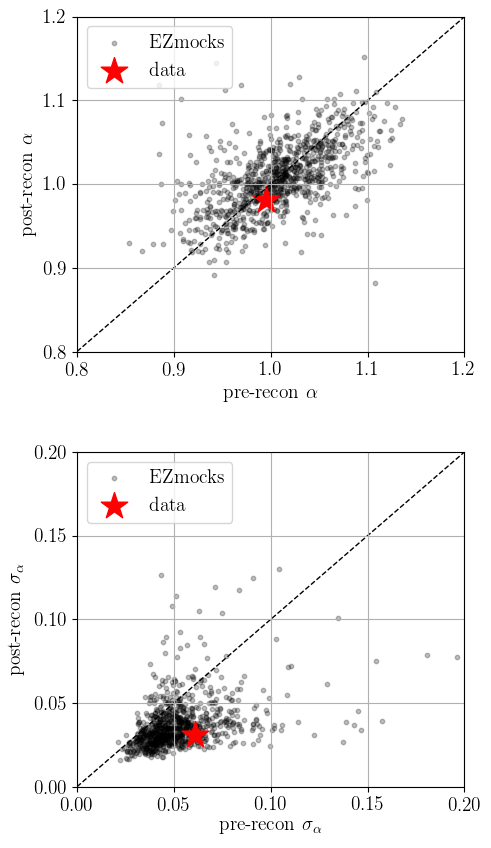} 
\caption{Comparison of the NGC+SGC pre- and post-reconstruction BAO fit results for the 1000 EZmocks (gray dots) and the data (red star). The top panel displays the $\alpha$ BAO parameter, and the bottom panel displays the uncertainty on $\alpha$.}
\label{fig:alpha_ezmock}
\end{center}
\end{figure}

The pre- and post-reconstruction fits on the individual EZmocks are displayed in Figure \ref{fig:alpha_ezmock}, and the results of the post-reconstruction fits to individual EZmock realisations are presented in Table \ref{tab:baomock}.
The fiducial case has `detections' (defined as having a $\Delta \chi^2 =1$ region within $0.8 < \alpha < 1.2$) for more that 96 percent of the realisations, but more than 10 percent of NGC/SGC individually do not have such detections. 
\citet{de-mattia20} find a similar fraction of no `detections' in the individual NGC/SGC when analysing the EZmocks in the Fourier space.
We find little gain is achieved by taking the mean result of the $\chi^2(\alpha)$ across the five bin centres.
For the ease of reproducibility and sharing/comparing results, we use will use bin centres with no shift (i.e., the first bin contains pairs with separation $0 < s < 5 h^{-1}$Mpc) as the fiducial result.

\begin{table}
\centering
\caption{Statistics for post-reconstruction BAO fits on the 1000 EZmocks. $\langle\alpha\rangle$ is the mean measured BAO parameter with $1\sigma$ bounds within the range $0.8 < \alpha < 1.2$. $\langle\sigma\rangle$ is the mean of the uncertainty obtained from $\Delta\chi^2=1$ region and $S$ is the standard deviation of these $\alpha$. $N_{\rm det}$ is the number of realisations with such $1\sigma$ bounds. The $\xi$ bin size is 5$h^{-1}$Mpc, unless noted otherwise. Tests of shifting bin centres are noted by $+x$, with $x$ representing the shift in $h^{-1}$Mpc. For these fits, we use damping parameters $\Sigma_{\perp,||} =  4,7h^{-1}$Mpc unless otherwise noted. Results labelled `combined' represent cases where the mean of the $\chi^2(\alpha)$ across five bin centres has been used.}
\begin{tabular}{lcccccc}
\hline
\hline
case (+bin shift)  & $\langle\alpha\rangle$ & $\langle\sigma\rangle$ & $S$ & $N_{\rm det}$ & $\langle\chi^2\rangle$/dof\\
\hline
EZmocks:\\
fiducial & 1.008 & 0.040 & 0.042 & 963 & 31.8/31\\
 +1  & 1.008 & 0.041 & 0.042 & 962 & 31.9/31\\
 +2  & 1.008 & 0.040 & 0.043 & 953 & 31.9/31\\
 +3  & 1.006 & 0.039 & 0.042 & 958 & 31.8/31\\
 +4  & 1.008 & 0.040 & 0.042 & 963 & 31.8/31\\
 combined & 1.008 & 0.040 & 0.041 & 961 & 31.9/31\\
 $\Delta s =8 h^{-1}$Mpc & 1.006 & 0.040 & 0.043 & 955 & 18.2/17\\
 $\Sigma_{\perp,||} = 3,5h^{-1}$Mpc & 1.008 & 0.038 & 0.042 & 965 & 31.9/31\\
NGC & 1.005 & 0.051 & 0.048 & 887 & 15.4/15\\
SGC & 1.006 & 0.054 & 0.054 & 861 & 15.4/15\\
\hline
\label{tab:baomock}
\end{tabular}
\end{table}

\subsection{BAO measurement from the DR16 ELG correlation function}

\begin{figure}
    \centering
    \includegraphics[width=0.45\textwidth]{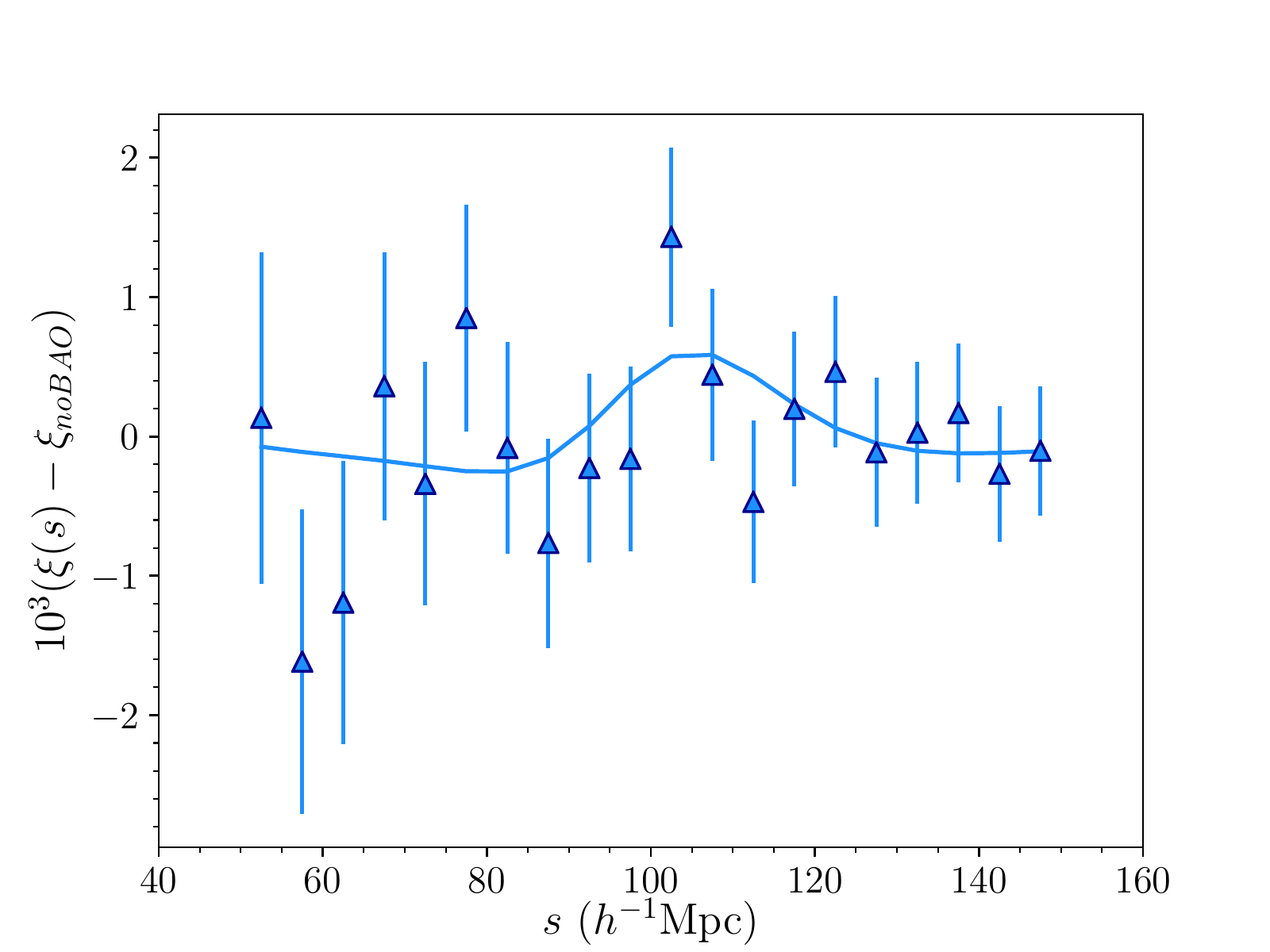}
    \caption{The NGC+SGC post-reconstruction correlation function compared to the best-fit model, both with the smooth component of the model subtracted. }
    \label{fig:xibao}
\end{figure}

\begin{figure}
    \centering
    \includegraphics[width=0.45\textwidth]{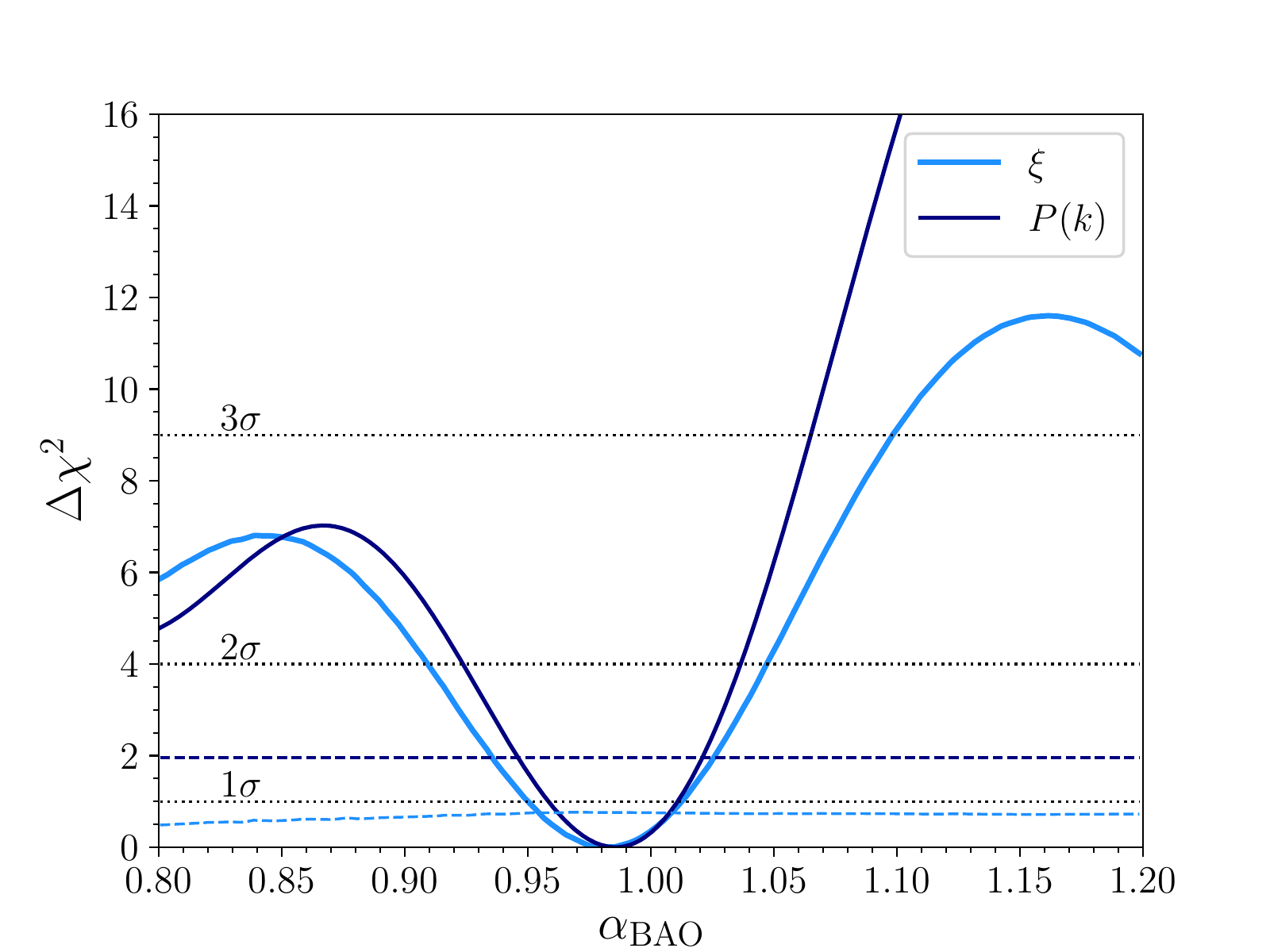}
    \caption{The BAO likelihood obtained from the combination of NGC and SGC results. We display our results ($\xi$) and also the Fourier-space ($P(k)$) results from \citet{de-mattia20}. The dashed curves show the results for the model with no BAO.}
    \label{fig:baolik}
\end{figure}

We use the post-reconstruction DR16 ELG correlation function to obtain a 3.2 percent measurement of $\frac{D_V(z_{\rm eff}=0.845)}{r_{\rm drag}} = 18.23\pm 0.58$.
This result is obtained from fitting the NGC and SGC results separately and adding their $\chi^2(\alpha)$.
This quoted result is a Gaussian approximation to the full likelihood; any cosmological tests should use the full non-Gaussian likelihood.
Our Gaussian approximation to the likelihood is to use the $\Delta\chi^2=1$ region as the 1$\sigma$ width.
The result is converted from $\alpha=0.981\pm0.031$ (Eq. ~\ref{eq:dvrd}).
The $\chi^2$/dof is slightly high, at 44.4/31, but a greater $\chi^2$ is expected 5.6 percent of the time under Gaussian expectations.

Figure \ref{fig:xibao} displays the result of our BAO fit. Here, we subtract the smooth, `no BAO' component of the best-fit model from both the data and the total best-fit model.
We display the inverse variance weighted mean of the NGC and SGC results.
The $\Delta\chi^2(\alpha)$ likelihood associated with this fit is displayed in Figure \ref{fig:baolik}, using a solid curve (labelled $\xi$).
It has a significant non-Gaussian component that becomes more pronounced far from the maximum likelihood.
Also shown is the $\Delta\chi^2(\alpha)$ when using a template with no BAO feature, using dashed curves.
There is only a mild ($\Delta\chi^2<1$) preference for the model with BAO. However, the no BAO model $\chi^2(\alpha)$ is nearly flat and has no local minima. Thus,
the precision of our result is produced by the fact that, while a smooth model is not a significantly worse fit to the data, a model {\it with} a BAO far from the maximum likelihood is a significantly worse fit to the data.

Figure \ref{fig:baolik} also displays the $\chi^2(\alpha)$ obtained from Fourier-space analysis in \citet{de-mattia20} (labelled $P(k)$).
The results of the two studies are clearly consistent in terms of the location of the BAO feature, but the $P(k)$ results are more precise.
The detailed tests presented in \citet{de-mattia20} demonstrate the robustness of their result and we thus recommend it is used for the DR16 ELG BAO measurement, given its increased precision.

We present a series of robustness test in Table \ref{tab:baodata}.
The most notable results from the table are those that show our measurements come almost entirely from the SGC data. This is not surprising given the $\xi_0$ displayed in Figure \ref{fig:xicompez}.
It is not particularly surprising that the NGC data does not provide a BAO measurement on its own: we find the same in more than 10 percent of the fits applied to the EZmocks.
This would happen somewhat less if the BAO signal in the EZmocks was consistent with our assumed $\Sigma_{\perp,||} = 3,5h^{-1}$Mpc.
Given 3.7 percent of the NGC+SGC fits to the EZmocks result in no BAO measurement, we believe it would remain at least a 5 percent probability.
Conversely, we are somewhat lucky with the SGC result, as 9.2 percent of the EZmocks have an uncertainty less than 0.033. This result would become more common if the EZmocks had a BAO signal consistent with $\Sigma_{\perp,||} = 3,5h^{-1}$Mpc.
This analysis suggests our results are not particularly unusual. While the NGC result does not afford a BAO measurement, we can use a Gaussian approximation and take the NGC+SGC and SGC only results to solve for the Gaussian equivalent of the impact of including the NGC result.
We find the NGC result is thus equivalent to $\alpha_{\rm NGC} = 0.91\pm0.10$; i.e., this result, added in quadrature with the SGC only result reproduces the NGC+SGC result.

As is typical for BAO measurements, the arbitrary choices in our analysis have a small effect on our measured $\alpha$.
Increasing the damping parameters to $\Sigma_{\perp,||} = 4,7h^{-1}$Mpc (from $3,5h^{-1}$Mpc) decreases $\alpha$ by $<0.1\sigma$ but does increase the estimated uncertainty by 16 percent.
Removing the prior on $B$ (which, in the fiducial modelling is a Gaussian prior of width 40 percent around the best-fit between $50 < s < 55h^{-1}$ Mpc) shifts the result higher by ~$\sigma/3$.
In this case, the NGC result prefers $B=0$ at all $\alpha$ and result comes entirely from the SGC.
The 10 percent decrease in the uncertainty comes from the fact that the $B$ value in the SGC can become greater than otherwise allowed and a stronger BAO feature is preferred in the SGC.
A 0.55$\sigma$ shift to a lower $\alpha$ value is observed when setting the polynomial terms to 0.
Once the number of polynomial terms is increased to at least two, the $\alpha$ result changes by less than 0.002.
The result is also stable to better than $0.1\sigma$ if we cut the sample to $z>0.7$, though doing so increases the uncertainty by 29 percent.
Finally, the uncertainty is decreased by nearly a factor of 2 via the application of reconstruction, but the $\alpha$ value shifts by less than the decrease in the uncertainty.
We conclude that, while there are puzzling aspects of the DR16 eBOSS ELG sample, the BAO measurements we extract from the sample are robust.

\begin{table}
\centering
\caption{Results for BAO fits to the DR16 ELG data. The fiducial $\xi$ case uses post-reconstruction data with 5$h^{-1}$Mpc bin size, centres in the range $50 < s < 150h^{-1}$Mpc, $\Sigma_{\perp,||} = 3,5h^{-1}$Mpc, and $0.6 < z < 1.1$.  }
\begin{tabular}{lcc}
\hline
\hline

{\bf Measurement}  & $\frac{D_V(z_{\rm eff}=0.845)}{r_{\rm drag}} = 18.23\pm 0.58$\\

\hline
{\bf Robustness tests}\\
case & $\alpha$ & $\chi^2$/dof\\
\hline
Post-recon. SGC+NGC:\\
fiducial & 0.981$\pm$0.031 & 44.4/31\\
$\Sigma_{\perp,||} = 4,7h^{-1}$Mpc & 0.979$\pm$0.036 & 44.5/31\\
no $B$ prior & 0.990$\pm$0.030 & 37.4/33\\
$A_n = 0$ & 0.964$\pm$0.035 & 51.8/37\\
$A_{1,2} = 0$ & 0.964$\pm$0.035 & 49.9/35\\
$A_{2} = 0$ & 0.980$\pm$0.033 & 47.6/33\\
$+A_{3} $ & 0.979$\pm$0.034 & 42.9/29\\
+1 & 0.978$\pm$0.033 & 50.1/31\\
+2 & 0.994$\pm$0.034 & 42.4/31\\
+3 & 0.985$\pm$0.031 & 39.4/31\\
+4 & 0.986$\pm$0.029 & 44.0/31\\
combined & 0.985$\pm$0.032 & 44.1/31\\
$P(k)$ \citep{de-mattia20} & 0.986$^{+0.025}_{-0.028}$ & --\\
Sample variations:\\
$z > 0.7$& 0.983$\pm$0.040 & 43.0/31\\
SGC & 0.989$\pm$0.033 & 17.2/15\\
NGC & no detection & 18.8/15\\
Pre-recon. & 0.995$\pm$0.061 & 40.2/31\\
\hline
\label{tab:baodata}
\end{tabular}
\end{table}

\section{Conclusion} \label{sec:conclusion}

We have presented the eBOSS/ELG DR16 spectroscopic data, the construction of the LSS catalogues, and the spherically averaged BAO analysis in configuration space.
The LSS catalogues are publicly available\footnote{A link to webpage will be provided after DR16 papers are accepted for publication.}, and used in two companions papers analysing the anisotropic clustering of the sample, \citet[][Fourier space]{de-mattia20} and \citet[][configuration space]{tamone20}.

After having described the observations of the 269,243 ELG spectra over 1170 deg$^2$, we detailed the $\zspec$ measurement procedure: thanks to pipeline improvements, the rate of redshift failures is decreased from 17 to 10 percent, while simultaneously decreasing the rate of catastrophic redshifts (from 0.5 to 0.3 percent), estimated from repeat observations and visual inspections.

We then described the construction of the LSS catalogues, which are required for the cosmological analyses.
Unlike other eBOSS tracers selected on SDSS imaging, the ELGs have been selected on a preliminary release of the DECaLS imaging; as a consequence the LSS construction requires a special attention.
For the data, we restrict to the 173,736 ELGs with a reliable $\zspec$ measurement with $0.6 < \zspec < 1.1$.
We extensively described the angular veto masks resulting from masking at the target selection step and \textit{a posteriori} masking for ensuring reliable galaxy observations.
We then defined the weights that correct for non-cosmological fluctuations; noticeably, the redshift failure correction accounts for the dependence on the observation conditions and on the instrumental patterns, which is significant due to the low SN of the ELG spectra.
Another feature specific to that ELG sample we need to correct for is the dependence of the redshift distribution with the imaging depth: shallow imaging regions tend to have more contamination from low-redshift objects entering the selection $grz$-box; we account for that effect with an \textit{ad hoc} method reproducing the effect in the randoms.

Lastly, we presented a spherically averaged BAO measurement on the reconstructed monopole.
The ELG data present a strong BAO feature in the SGC and no significant BAO feature in the NGC; analysing 1000 approximate EZmocks suggests that this result is not particularly unusual.
When combining the SGC and the NGC, the data has a feature consistent with that of the BAO, providing a 3.2 percent measurement of $D_V(z_{\rm eff}=0.845)/r_{\rm drag} = 18.23\pm 0.58$. 

The analysis presented in this paper, along with the ones presented in \citet{de-mattia20} and \citet{tamone20} are likely to provide valuable tools in the ELG clustering analysis, paving the way for next generation massive BAO surveys, which will mostly target ELGs, as DESI, PFS, \textit{Euclid}, or \textit{WFIRST}.

\section*{Acknowledgments}

AR and JPK acknowledge support from the ERC advanced grant LIDA.
AR, CZ, and AT acknowledge support from the SNF grant 200020\_175751.
AdM acknowledges support from the P2IO LabEx (ANR-10-LABX-0038) in the framework "Investissements d’Avenir" (ANR-11-IDEX-0003-01) managed by the Agence Nationale de la Recherche (ANR, France).
AJR is grateful for support from the Ohio State University Center for Cosmology and Particle Physics.
Authors acknowledge support from the ANR eBOSS project (ANR-16-CE31-0021) of the French National Research Agency.
S. Alam is supported by the European Research Council through the COSFORM Research Grant (\#670193).
S. Avila was supported by the MICUES project, funded by the European Union’s Horizon 2020 research programme under the Marie Sklodowska-Curie Grant Agreement No. 713366 (InterTalentum UAM).
ADM was supported by the U.S. Department of Energy, Office of Science, Office of High Energy Physics, under Award Number DE-SC0019022.
JM gratefully acknowledges support from the U.S. Department of
Energy, Office of Science, Office of High Energy Physics under Award
Number DE-SC002008 and from the National Science Foundation under
grant AST-1616414.
EMM acknowledges support from the European Research Council (ERC) under the European Union’s Horizon 2020 research and innovation programme (grant agreement No 693024).
VGP acknowledges support from the European Union's Horizon 2020 research and innovation programme (ERC grant \#769130).
GR acknowledges support from the National Research Foundation of Korea (NRF) through Grants No. 2017R1E1A1A01077508 and No. 2020R1A2C1005655 funded by the Korean Ministry of Education, Science and Technology (MoEST), and from the faculty research fund of Sejong University.

Funding for the Sloan Digital Sky Survey IV has been provided by the Alfred P. Sloan Foundation, the U.S. Department of Energy Office of Science, and the Participating Institutions. SDSS acknowledges support and resources from the Center for High-Performance Computing at the University of Utah. The SDSS web site is www.sdss.org.

SDSS is managed by the Astrophysical Research Consortium for the Participating Institutions of the SDSS Collaboration including the Brazilian Participation Group, the Carnegie Institution for Science, Carnegie Mellon University, the Chilean Participation Group, the French Participation Group, Harvard-Smithsonian Center for Astrophysics, Instituto de Astrofísica de Canarias, The Johns Hopkins University, Kavli Institute for the Physics and Mathematics of the Universe (IPMU) / University of Tokyo, the Korean Participation Group, Lawrence Berkeley National Laboratory, Leibniz Institut für Astrophysik Potsdam (AIP), Max-Planck-Institut für Astronomie (MPIA Heidelberg), Max-Planck-Institut für Astrophysik (MPA Garching), Max-Planck-Institut für Extraterrestrische Physik (MPE), National Astronomical Observatories of China, New Mexico State University, New York University, University of Notre Dame, Observatório Nacional / MCTI, The Ohio State University, Pennsylvania State University, Shanghai Astronomical Observatory, United Kingdom Participation Group, Universidad Nacional Autónoma de México, University of Arizona, University of Colorado Boulder, University of Oxford, University of Portsmouth, University of Utah, University of Virginia, University of Washington, University of Wisconsin, Vanderbilt University, and Yale University.

This paper presents observations obtained at Cerro Tololo Inter-American Observatory, National Optical Astronomy Observatory (NOAO Prop. ID: 2014B-0404; co-PIs: D. J. Schlegel and A. Dey), which is operated by the Association of Universities for Research in Astronomy (AURA) under a co\"operative agreement with the National Science Foundation. This paper also includes DECam observations obtained as part of other projects, namely 
the Dark Energy Survey (DES, NOAO Prop. ID: 2012B-0001).

DECaLS used data obtained with the Dark Energy Camera (DECam), which was constructed by the Dark Energy Survey (DES) collaboration. Funding for the DES Projects has been provided by the U.S. Department of Energy, the U.S. National Science Foundation, the Ministry of Science and Education of Spain, the Science and Technology Facilities Council of the United Kingdom, the Higher Education Funding Council for England, the National Center for Supercomputing Applications at the University of Illinois at Urbana-Champaign, the Kavli Institute of Cosmological Physics at the University of Chicago, Center for Cosmology and Astro-Particle Physics at the Ohio State University, the Mitchell Institute for Fundamental Physics and Astronomy at Texas A\&M University, Financiadora de Estudos e Projetos, Funda{\c c}{\~a}o Carlos Chagas Filho de Amparo, Financiadora de Estudos e Projetos, Funda{\c c}{\~a}o Carlos Chagas Filho de Amparo {\`a} Pesquisa do Estado do Rio de Janeiro, 
Conselho Nacional de Desenvolvimento Cient{\'i}fico e Tecnol{\'o}gico and the Minist{\'e}rio da Ci{\^e}ncia, Tecnologia e Inovac{\~a}o, the Deutsche Forschungsgemeinschaft and the Collaborating Institutions in the Dark Energy Survey. 
The Collaborating Institutions are Argonne National Laboratory, the University of California at Santa Cruz, the University of Cambridge, 
Centro de Investigaciones En{\'e}rgeticas, Medioambientales y Tecnol{\'o}gicas-Madrid, the University of Chicago, University College London, the DES-Brazil Consortium, the University of Edinburgh, 
the Eidgen{\"o}ssische Technische Hoch\-schule (ETH) Z{\"u}rich, 
Fermi National Accelerator Laboratory, the University of Illinois at Urbana-Champaign, 
the Institut de Ci{\`e}ncies de l'Espai (IEEC/CSIC), 
the Institut de F{\'i}sica d'Altes Energies, 
Lawrence Berkeley National Laboratory, 
the Ludwig-Maximilians Universit{\"a}t M{\"u}nchen and the associated Excellence Cluster Universe, 
the University of Michigan, the National Optical Astronomy Observatory, the University of Nottingham, the Ohio State University, the University of Pennsylvania, the University of Portsmouth, SLAC National Accelerator Laboratory, Stanford University, the University of Sussex, and Texas A\&M University. 

This research used resources of the National Energy Research Scientific Computing Center, a DOE Office of Science User Facility supported by the Office of Science of the U.S. Department of Energy under Contract No. DE-AC02-05CH11231.

This work used resources from the Sciama High Performance Computing cluster, which is supported by the Institute of Cosmology and Gravitation and the University of Portsmouth. 

\textit{Authors contribution.}
AR led this paper, the supervising of the spectroscopic data acquisition, the generation of intermediate catalogues from the pipeline outputs, the validation of the \texttt{redrock} $\zspec$ measurements, the building of the veto masks and of the $w_{\rm sys}$ and $w_{\rm noz}$ weights.
AdM led the generation of the LSS catalogues from intermediate catalogues, the implementation of systematics in the mocks, the implementation of the $n(z)$-depth dependence, and the building of the $w_{\rm cp}$ and $w_{\rm FKP}$ weights.
AJR led the BAO fitting.
CZ led the EZmocks realisation.
JB, KSD, and HdMdB produced the \texttt{redrock} ELG $\zspec$ measurements.
Other co-authors provided valuable input products or feedback for the analysis.

\bibliographystyle{mnras}
\bibliography{ms}


\bsp	
\label{lastpage}
\end{document}